\documentclass[10pt, letterpaper, notitlepage, oneside]{article}

\usepackage[
	letterpaper,
	bottom      = 1.00in,
	left        = 0.75in,
	right       = 0.75in,
	top         = 1.00in
]{geometry}
\usepackage[utf8]{inputenc}
\usepackage[nopar]{lipsum}

\usepackage{fancyhdr}
\fancypagestyle{specialfooter}{
	\fancyhf{}
	\lhead{Published in \textit{Physics of Fluids}: \href{https://doi.org/10.1063/1.4946758}{\url{https://doi.org/10.1063/1.4946758}} \hfill \thepage}
	\cfoot{$\dag$ \textbf{Email address for correspondence:} g\_dilabb{\MVAt}encs.concordia.ca}
}

\usepackage{hyperref}
\hypersetup{
	breaklinks = true,
	citecolor  = blue,
	colorlinks = true,
	filecolor  = blue,
	linkcolor  = blue,      
	urlcolor   = blue
}

\usepackage{titlesec}
\titleformat*{\section}{\Large\bfseries}
\titleformat*{\subsection}{\normalsize\bfseries\filcenter}
\titleformat*{\subsubsection}{\normalsize\bfseries}
\renewcommand{\thesection}{\Roman{section}}

\makeatletter
\renewcommand\@seccntformat[1]{\csname the#1\endcsname.\quad}
\makeatother

\usepackage{amsfonts}
\usepackage{amsmath}
\usepackage{marvosym}
\usepackage{textcomp}
\usepackage{upgreek}
\newcommand{\volume}{{\ooalign{\hfil$V$\hfil\cr\kern0.08em--\hfil\cr}}}

\usepackage{caption}
\usepackage{cleveref}
\usepackage{graphicx}
\usepackage{epstopdf}
\usepackage[
	font = normalsize
]{subfig}
\usepackage{xparse}
\crefformat{subfigure}{#2\onlyletter{#1}#3}
\crefrangeformat{subfigure}{#1--#3\onlyletter{#2}#4}
\ExplSyntaxOn
\NewDocumentCommand{\onlyletter}{m}{
	\tl_set:Nx  \l_tmpa_tl { #1 }
	\tl_item:Nn \l_tmpa_tl { -1 }
}
\ExplSyntaxOff

\usepackage{arydshln} 
\usepackage{multirow}
\usepackage{setspace}

\usepackage{apacite} 
\let\cite\shortcite
\let\citeA\shortciteA

\newcommand{\noop}[1]{}

\begin{document}
	
	\thispagestyle{specialfooter}
	
	\noindent
	\textbf{\LARGE Linear and weakly nonlinear global instability of a fluid flow\\\\ through a collapsible channel}\hfill\break
	
	\begin{changemargin}{0.25in}{0.00in}
		{\large Amaouche M, Di Labbio G$^{\dag}$}\hfill\break
		\textit{Laboratory of Cardiovascular Fluid Dynamics, Concordia University, Montr\'{e}al, QC, Canada, H3G 1M8}\hfill\break
		
		Interactions between an internal flow and wall deformation occur in many biological systems. Such interactions can involve a complex and rich dynamical behavior and a number of peculiarities which depend on the flow parameter range. The aim of this paper is to present a variant (obtained via a weighted residual approach) of the averaged one-dimensional model derived by Stewart \textit{et al}.\ ``Local and global instabilities of flow in a flexible-walled channel,'' Eur.\ J.\ Mech.\ B/Fluids \textbf{28}, 541-557 (2009)]. The asymptotic expansions for small Reynolds numbers of these two models, compared to the exact solution obtained from the lubrication approach, reveal some quantitative difference, even at higher Reynolds numbers. Qualitatively, the two models give similar results at least at a linear level. It is shown that for relatively low membrane tension ($T$), there are distinct regions in the $(T,R)$ parameter space where steady bifurcating flows may occur. These flows can also be observed at vanishingly small Reynolds numbers combined with relatively high membrane tension. At sufficiently high $T$ and $R$, the bifurcating flow is rather time periodic. A weakly nonlinear analysis is then performed in both cases leading to the derivation of evolution equations for the amplitudes of the bifurcating flows. The amplitude equations show that the saddle node bifurcation has a transcritical character while the Hopf bifurcation is either subcritical or supercritical, depending both on the mode number and membrane tension.
		
		\hfill\break
		* \textit{Data pertaining to this article will be made available from the authors upon reasonable request.}\hfill\break
		* \textit{The authors have no conflicts of interest to declare.}\hfill\break
		* \textit{Please cite as}: Amaouche, M., \& Di Labbio, G. (2016). Linear and weakly nonlinear global instability of a fluid flow through a collapsible channel. \textit{Physics of Fluids}, \textit{28}(4), 044106.
		
		\hfill\break
		\raisebox{0.95pt}{\small\textcopyright} 2016 American Institute of Physics Publishing. This manuscript version is made available under the AIP Publishing LLC Transfer of Copyright Agreement, more information regarding usage terms can be found at \href{https://publishing.aip.org/resources/researchers/rights-and-permissions/permissions/}{\url{https://publishing.aip.org/resources/researchers/rights-and-permissions/permissions/}}. This article may be downloaded for personal use only. Any other use requires prior permission of the author and AIP Publishing. This article appeared in ``Amaouche, M., \& Di Labbio, G. (2016). Linear and weakly nonlinear global instability of a fluid flow through a collapsible channel. \textit{Physics of Fluids}, \textit{28}(4), 044106.'' and may be found at \href{https://doi.org/10.1063/1.4946758}{\url{https://doi.org/10.1063/1.4946758}}.
	\end{changemargin}
	
	\section{\label{sec:Intro}Introduction}
	
		We investigate linear and weakly nonlinear instability of a viscous and incompressible Newtonian fluid driven by a prescribed pressure gradient through a finite length plane channel for which a segment of one wall is a membrane under longitudinal tension. This hydroelastic system is the plane schematic representation of the Starling resistor, first introduced by \citeA{Pedley92}, in which a segment of flexible tubing is mounted between two rigid tubes and enclosed in a pressure chamber. The latter allows the external pressure acting on the flexible part of the tube to be controlled independently of the internal fluid pressure. When the elastic tube is subjected to a sufficiently large compression, it buckles asymmetrically along part of its length, leading to an increase of its compliance, thereby becoming potentially capable to develop large amplitude self-excited oscillations when the internal flow rate exceeds a certain critical value \cite{Bertram03}.
		
		The propensity of flexible tubes to exhibit significant deformation has motivated a large number of investigations because of its relevance in a wide variety of physiological phenomena. For reviews of pertinent physiological applications, we refer to \citeA{Grotberg04}. These include noise-generating instabilities such as Korotkoff sounds produced during blood pressure measurement in the compressed brachial artery, snoring during inhalation, and flow limitation in lung airways such as wheezing during expiration. Self-excited oscillations are responsible for the generation of speech in the human larynx and bird song in the avian syrinx. It is presently known that self-excited oscillations in the Starling resistor arise in multiple distinct modes whose interactions can lead to a complex and rich dynamical behavior of the subsequent flow. Substantial advances have been achieved in understanding the mechanisms underlying these global instabilities, which are intimately related to the boundary conditions at the upstream and downstream ends of the flexible membrane.
		
		Studies about global instabilities in finite-length asymmetric channels typically model the compliant wall as a tensioned membrane and neglect properties such as wall inertia, wall elasticity, and bending stiffness. The pressure is typically held fixed far downstream and either the flow rate or the pressure may be prescribed upstream. In either case, the system may exhibit a wide variety of self-excited oscillations. \citeA{Jensen03} identified, using high Reynolds number asymptotics, a mechanism for the generation of long wavelength and high frequency mode $1$ sloshing oscillations when the membrane is under high tension. They provided an explanation of this instability mechanism in terms of the fluid's energy balance and found that this high frequency mode requires the upstream pressure to be prescribed and the upstream rigid channel segment to be shorter than that downstream. In these conditions, the in and out motion of the membrane forces significant oscillatory volume flow in the rigid parts of the channel and so energy can be transferred from the inflow to the oscillations. This high frequency mode has also been captured by the integral model (IM) derived by \citeA{Stewart09} who found that the base flow exhibits static and oscillatory global instabilities and showed how self-excited oscillations can arise entirely from wave reflections from the rigid parts of the system. Hence, an unstable global mode is composed of interacting local modes, each of which can be stable, with instability being driven primarily by the boundary conditions. Locally unstable modes are naturally expected to also contribute to the occurrence of global instability. Analyzing three different systems, \citeA{Doare06} have proposed a method to quantify the contributions of local instability and of wave reflection in the global instability of one-dimensional systems.
		
		While quantitative accuracy can be guaranteed only for sufficiently small Reynolds numbers by the IM, the predictions for larger Reynolds numbers were shown to capture many important features computed by direct numerical simulation of the basic equations and by sophistical asymptotic analysis \cite{Stewart10}. The model derived by \citeA{Stewart09} was also employed by \citeA{Xu13, Xu14} to analyze the low frequency mode $1$ oscillations in the case of a fixed inflow rate. Changing the coefficient of convective inertia, they showed that the model's predictions are qualitatively insensitive to the velocity profile assumption. Their weakly nonlinear analyses show in particular more than one type of unsteady behavior that can occur in the neighborhood of a particular organizing point. The results indicate that two possible states of sustained oscillations can arise, depending on whether the length of the downstream rigid segment is of the same order or much greater than the membrane length. For a detailed review of one-dimensional models we refer to \citeA{Pedley15}.
		
		As emphasized by many authors, the IM is a tractable model which performs well from small to moderate Reynolds numbers. However, despite its success in the nonlinear regime, this model does not incorporate the effects of inertia through the parabolic velocity assumption. Corrections to the parabolic velocity profile in order to coherently account for the inertial effects are therefore required for an accurate prediction of the linear instability threshold, at least for small inertia. Our purpose here is to obtain such a coherent first order model which also performs well for moderate Reynolds numbers as the IM does.
		
		This paper is organized as follows. In Section \ref{sec:Prob}, we follow \citeA{Stewart09} to present the physical problem and its governing equations. In Section \ref{sec:Model}, we derive the weighted integral model (WIM) and compare its asymptotic expansion with that obtained from the IM and the exact lubrication solution. Linear and weakly nonlinear stability of the base flow are examined in Sections \ref{sec:LinStab} and \ref{sec:WeakNL}, respectively. Representative numerical results from the two models are then given and compared in Section \ref{sec:Num}. Finally, a conclusion is offered in Sec.\ \ref{sec:Conc}.
		
	\section{\label{sec:Prob}Physical Problem and Basic Equations}
		
\begin{figure}[!h]
			\centering
			\includegraphics[width=0.8\textwidth]{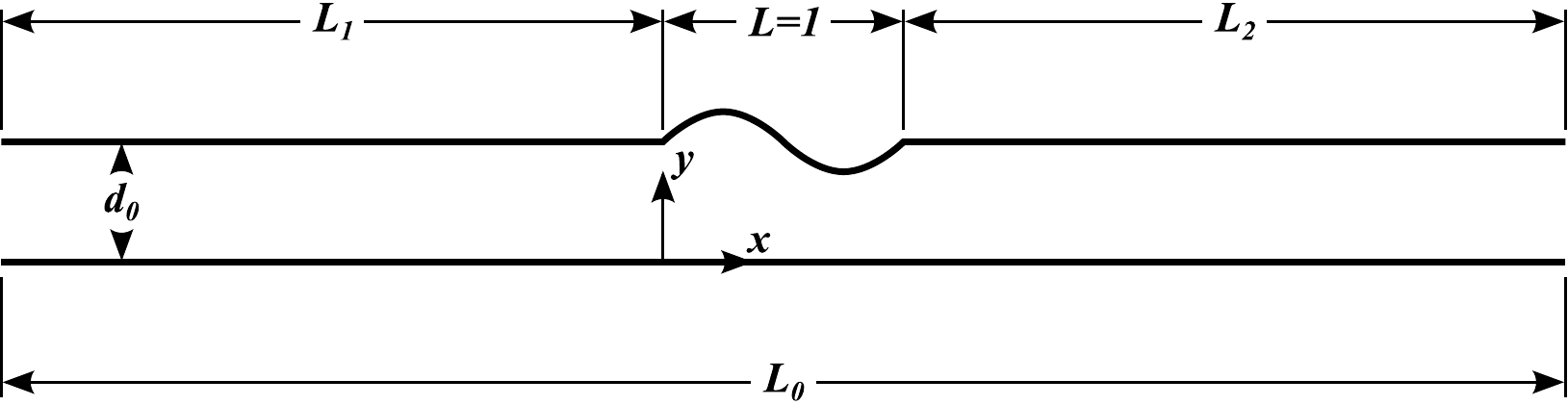}
		\caption{Schematic representation of the problem of Poiseuille flow in an asymmetric channel with a tensioned membrane fixed in one wall.}
		\label{fig:ProbDef}
		\end{figure}

		The physical system consists of a Newtonian incompressible fluid with density $\rho^*$ and dynamic viscosity $\mu^*$ flowing in a slender planar channel of total length $L^*_0$ (the asterix is used to denote dimensional quantities). A segment of one wall of the channel is made of a thin elastic membrane, of length $L^*$ and longitudinal tension $T^*$, mounted between two rigid segments with lengths $L^*_1$ upstream and $L^*_2$ downstream (see Fig.\ \ref{fig:ProbDef} for a schematic of the problem). To simplify matters, the other properties of the membrane (inertia, damping, and bending stiffness) are neglected. When the membrane is flat, the channel has a uniform width $d^*_0$ which is assumed much smaller than $L^*$. This introduces a small perturbation parameter $\epsilon = d^*_0/L^*$ which allows for appreciable simplifications in the formulation. The flow is driven by a prescribed constant pressure gradient between the channel exit where the pressure is $p^* = 0$ and the channel entrance where $p^* = p^*_u$. We introduce a Cartesian coordinate system with the $x^*$ axis lying on the rigid wall such that $x^*$ increases downstream and $x^* = 0$ corresponds to the abscissa of the upstream end of the membrane. The $y^*$ axis is oriented inside the flow. The smallness of $\epsilon$ implies a separation of scales between $x^*$ and $y^*$ which in turn dictates a different treatment of the streamwise and cross-stream momentum equations as in boundary layer theory. Hence, making use of a suitable scaling which will be defined shortly and neglecting terms of $O(\epsilon^2)$ and higher, the dimensionless boundary layer type equations read
		\begin{gather}
			u_x + v_y = 0, \label{eq:cont}\\
			R(u_t + uu_x + vu_y + p_x) = u_{yy}, \label{eq:x-mom}\\
			p_y = 0, \label{eq:y-mom}
		\end{gather}
		where $u$ and $v$ are the streamwise and cross-stream velocity components, $t$ is time, and $R = U^*_0{d^*_0}^2/L^*{\nu}^*$ is the Reynolds number; ${\nu}^*$ being the kinematic viscosity and $U^*_0 = p^*_u{d^*_0}^2/12{\mu}^*L^*_0$ being the averaged velocity of the base flow. Equations (\ref{eq:cont})-(\ref{eq:y-mom}) are rendered dimensionless by scaling $x^*$, $y^*$, $t^*$, $u^*$, $v^*$, and $p^*$ with $L^*$, $d^*_0$, $L^*/U^*_0$, $U^*_0$, $U^*_0d^*_0/L^*$, and ${\rho}^*{U^*_0}^2$, respectively. The flow is subject to the no-slip and no-penetration boundary conditions at the walls and to the normal stress balance which determines the shape $y = h(x,t)$ of the membrane, namely,
		\begin{gather}
			(u,v) = (0,0) \quad \text{at} \quad \left \lbrace
			\begin{array}{l}
				y = 0, \ -L_1 < x < 1 + L_2 \\
				y = 1, \  -L_1 < x < 0 \ \text{and} \ 1 < x < 1 + L_2
			\end{array},\right. \label{eq:bc-wall}\\
			(u,v,p) = (0,h_t,p_e - Th_{xx}) \quad y = h(x,t), \ 0 < x < 1, \label{eq:bc_mem}
		\end{gather}
		with $T = d_0^*T^*/{\rho}^*{U^*_0}^2{L^*}^2$ being the dimensionless tension parameter and $p_e(x)$ a prescribed external pressure defined below. When the membrane is flat, one obtains the base flow with velocity $(u_0,v_0) = (6y(1-y),0)$ from the equilibrium of viscous forces and the prescribed constant pressure gradient. The internal pressure distribution is dependent only on $x$ owing to (\ref{eq:y-mom}) and is given by $p(x) = p_u - p_u(x + L_1)/L_0$ with $p_u/L_0 = 12/R$. For there to be no pressure difference on either side of the membrane when it is flat, the same pressure distribution is imposed outside, namely,
		\begin{equation}
		\label{eq:ext_press}
			p_e(x) = p_u - \frac{p_u}{L_0}(x + L_1).
		\end{equation}
		
		Equation (\ref{eq:x-mom}) subject to the dynamic boundary condition at the membrane combined with (\ref{eq:ext_press}) provides the pressure distribution in the channel for $0 < x < 1$,
		\begin{equation}
		\label{eq:press}
			p = p_u - \frac{12}{R}(x + L_1) - Th_{xx}
		\end{equation}
		
		The momentum equation then becomes (for $0 < x < 1$):
		\begin{equation}
		\label{eq:mem_mom}
			R(u_t + uu_x + vu_y - Th_{xxx})= 12 + u_{yy}
		\end{equation}
		
	\section{\label{sec:Model}Model}
		
		Integrating the continuity equation over the channel width and accounting for the kinematic condition on the membrane yields the global equation
		\begin{equation}
		\label{eq:Icont}
			q_x + h_t = 0,
		\end{equation}
		where $q$ is the local flow rate. This shows that the flow rate, like the shape of the membrane, is a relevant variable. Consequently, it appears quite natural to seek another equation with suitable boundary conditions to obtain a closed system for these two genuine variables. Such a system was first derived by \citeA{Stewart09} for the Starling resistor. The IM is obtained by performing direct integration of the momentum equation and using a parabolic velocity profile resulting from the balance between viscous forces and the prescribed pressure gradient. Assuming such a velocity profile amounts to neglecting the distortion induced by inertia. For small Reynolds numbers, a suitable velocity profile accounting for the inertia effects would be of the form
		\begin{equation}
		\label{eq:uprof}
			u(x,y,t) = \frac{q(x,t)}{\displaystyle\int_0^{h(x,t)} u_0\left(\frac{y}{h}\right) \mathrm{d}y}u_0\left(\frac{y}{h}\right) + Ru_1(x,y,t) + O(R^2),
		\end{equation}
		where $u_0(y/h) = 6(y/h)(1 - y/h)$ and the correction  $u_1$ is such that $\int_0^h u_1 \mathrm{d}y = 0$. In a calculation up to $O(R)$, the correction $u_1$ is not needed to express inertial terms which are already of $O(R)$, but must be taken into account in the viscous term for consistency. Hence, up to $O(R)$ the averaged momentum equation would read
		\begin{equation}
		\label{eq:ImomJ}
			R\left(q_t - \frac{6}{5}\frac{q^2}{h^2}h_x + \frac{12}{5}\frac{q}{h}q_x - Thh_{xxx}\right) = 12\left(h - \frac{q}{h^2}\right) + Ru_{1y}\Big{|}_0^h.
		\end{equation}
		
		This is the IM derived by \citeA{Stewart09} complemented by the unknown quantity $Ru_{1y}\Big{|}_0^h$. The question then is what global procedure allows for the incorporation of $u_1$ without calculating it explicitly. This can be achieved by introducing a weighting function $f(y/h)$ before integrating the momentum equation. Hence, after two successive integrations by parts and using the no-slip condition at the walls, one obtains
		$$
			R\int_0^h f\left(\frac{y}{h}\right)\left\lbrace \tilde{u}_{0t} + \tilde{u}_0\tilde{u}_{0x} + \tilde{v}_0\tilde{u}_{0y} - Th_{xxx} \right\rbrace \mathrm{d}y = 12\int_0^h f\mathrm{d}y + \int_0^h uf_{yy} \mathrm{d}y + fu_y\Big{|}_0^h,
		$$
		where the tilde refers to the leading order term of $u$ in (\ref{eq:uprof}). One can see that the viscous term can be rendered free from $u_1$ if the weighting function is chosen so that $f_{yy} = K(x,t)$ and $f(0) = f(h) = 0$, $K(x,t)$ being an arbitrary function of $x$ and $t$. So $f$ is, apart from a multiplicative constant, nothing more than the base velocity $u_0(y/h)$.
		
		Further, the function $K(x,t)$ is chosen in order to have the same viscous term as in Equation (\ref{eq:ImomJ}); i.e., $K(x,t) = -12/h^2$ and therefore $f = 6(y/h)(1 - y/h)$. The integrated weighted momentum equation then reads
		\begin{equation}
		\label{eq:Imom}
			R\left({\gamma}q_t - \frac{6}{5}{\alpha}\frac{q^2}{h^2}h_x + \frac{12}{5}{\beta}\frac{q}{h}q_x - Thh_{xxx}\right) = 12\left(h - \frac{q}{h^2}\right),
		\end{equation}
		with $(\alpha,\beta,\gamma) = (1,1,1)$ for the IM and $(\alpha,\beta,\gamma) = (9/7,17/14,6/5)$ for the present model referred to as the WIM. It is to be noted (i) that the term $-12q/h^2$ arising from the weighted integration of the viscous force is exact in the sense that it is obtained without any assumption on the velocity profile, and (ii) up to $O(R)$, Equation (\ref{eq:Imom}) is the same as Equation (\ref{eq:ImomJ}) without the term $Ru_{1y}\Big{|}_0^h$ which is related to the unknown correction $u_1$.
		
		Now, following \citeA{Stewart09}, we now derive the boundary conditions associated to Equation (\ref{eq:Imom}) when the pressures at the ends of the tube are prescribed. Where the channel has rigid walls, Equation (\ref{eq:Imom}) can be simplified since $h = 1$, which yields $q_x = 0$ and therefore $q = q(t)$. Then, one is led to
		\begin{equation}
		\label{eq:press_wall}
			{\gamma}q_t = -p_x - \frac{12}{R}q.
		\end{equation}
		
		Integrating this equation subject to the prescribed pressures at the entrance and the exit of the channel yields
		\begin{gather}
			p\Big{|}_{x=0} = p_u - \left({\gamma}q_t + \frac{12}{R}q\right)L_1, \label{eq:press_0}\\
			p\Big{|}_{x=1} = \left({\gamma}q_t + \frac{12}{R}q\right)L_2. \label{eq:press_1}
		\end{gather}
		
		Combining Equations (\ref{eq:press_0}) and (\ref{eq:press_1}) with (\ref{eq:press}), one obtains the boundary conditions for the two variable system given by Equations (\ref{eq:Icont}) and (\ref{eq:Imom}),
		\begin{equation}
		\label{eq:Ibcs}
			\begin{array}{lll}
				x = 0: & h = 1, & Th_{xx} = L_1\left(\displaystyle\frac{12}{R}(q - 1) + {\gamma}q_t\right), \\
				x = 1: & h = 1, & Th_{xx} = -L_2\left(\displaystyle\frac{12}{R}(q - 1) + {\gamma}q_t\right).
			\end{array}
		\end{equation}
		
		In order to give evidence for the difference in predicting criticality for small Reynolds numbers between the WIM and the IM, we compare their asymptotic expansion as $R \rightarrow 0$ with the asymptotic solution of the full problem given by Equations (\ref{eq:cont}) and (\ref{eq:x-mom}). Using Equation(\ref{eq:Imom}) with $q$ being expanded in asymptotic series in $R$ (i.e., $q = q_0 + Rq_1 +\ \cdots$) and equating like powers of $R$, one obtains for the first two orders
		\begin{gather}
			q_0 = h^3, \label{eq:AE-WIM_q0}\\
			q_1 = Rh^3\left\lbrace \frac{T}{12}h_{xxx} + \left(\frac{3}{4}\gamma + \frac{1}{10}\alpha - \frac{3}{5}\beta\right)h^3h_x\right\rbrace.\label{eq:AE-WIM_q1}
		\end{gather}
		
		A \citeA{Benney66} type equation for the membrane shape is then derived from the continuity equation as
		$$
			h_t + (q_0 + Rq_1)_x = 0.
		$$
		
		In the range of Reynolds numbers where the Benney equation applies, the dynamics of the flow is slaved to its kinematics; i.e., the flow rate is adiabatically slaved to the membrane shape and it varies with $x$ and $t$ only through the dependence of the membrane shape on these variables.
		
		Performing a similar asymptotic expansion for the velocity in (\ref{eq:cont}) and (\ref{eq:x-mom}) yields
		\begin{gather}
			u_0 = 6y(h - y), \label{eq:AE-Base_u0}\\
			u_1 = R\left\lbrace \frac{3}{2}hh_x(h^3y - 2hy^3 + y^4) + \frac{T}{2}h_{xxx}(hy - y^2)\right\rbrace, \label{eq:AE-Base_u1}
		\end{gather}
		which correspond to
		\begin{gather}
			q_0 = h^3, \label{eq:AE-Base_q0}\\
			q_1 = Rh^3\left(\frac{3}{10}h^3h_x + \frac{T}{12}h_{xxx}\right). \label{eq:AE-Base_q1}
		\end{gather}
		
		Setting $(\alpha,\beta,\gamma) = (9/7,17/14,6/5)$ in (\ref{eq:AE-WIM_q1}), we retrieve the same correction for $q_1$ as that given by (\ref{eq:AE-Base_q1}). It is worth noting that the lubrication approach requires the explicit calculation of the first order correction of the velocity profile whereas the WIM does not. In spite of this, the two approaches yield the same results for the correction $q_1$, at least up to $O(R)$, while the IM provides a slightly different expression, namely,
		$$
			q_1 = Rh^3\left(\frac{1}{4}h^3h_x + \frac{T}{12}h_{xxx}\right).
		$$
		
	\section{\label{sec:LinStab}Linear Stability}
		
		Following common practice, we consider linear stability of the basic solution $(h,q) = (1,1)$ by setting $(h,q) = (1,1) + (\tilde{h},\tilde{q})$, where $(\tilde{h},\tilde{q})$ are small perturbations satisfying the following linear system:
		\begin{gather}
			\tilde{q}_x + \tilde{h}_t = 0, \label{eq:Pcont}\\
			R\left(\gamma\tilde{q}_t - \frac{6}{5}\alpha\tilde{h}_x + \frac{12}{5}\beta\tilde{q}_x - T\tilde{h}_{xxx}\right) = 12(3\tilde{h} - \tilde{q}), \label{eq:Pmom}\\
			\begin{array}{lll}
				x = 0: & \tilde{h} = 0, & T\tilde{h}_{xx} = L_1\left(\displaystyle\frac{12}{R}\tilde{q} + {\gamma}\tilde{q}_t\right), \\
				x = 1: & \tilde{h} = 0, & T\tilde{h}_{xx} = -L_2\left(\displaystyle\frac{12}{R}\tilde{q} + {\gamma}\tilde{q}_t\right).
			\end{array} \label{eq:Pbcs}
		\end{gather}
		
		The linearity of the system given by Equations (\ref{eq:Pcont})-(\ref{eq:Pbcs}) permits solutions proportional to $e^{{\sigma}t + {\lambda}x}$. The fixed point $(\tilde{h},\tilde{q}) = (0,0)$ may lose its stability either in favor of another fixed point via a saddle node bifurcation or in favor of a limit cycle via a Hopf bifurcation, depending on the parameter values. The former occurs when a real eigenvalue $\sigma$ crosses the imaginary axis ($\sigma = 0$) and the second corresponds to a pair of complex conjugate eigenvalues crossing.
		
		\subsection{\label{sec:SSBif}Steady-state bifurcation}
		
			In this case, a secondary fixed point bifurcates from the basic state. This corresponds to a steady flow governed by equations which can be derived from (\ref{eq:Pcont}) to (\ref{eq:Pbcs}) by ignoring the time dependence. Consequently, $\tilde{q}$ is an arbitrary constant, say $q_1^{(s)}$, and $\tilde{h}$ satisfies
			\begin{gather}
				\mathcal{L}_s\tilde{h} = \frac{12}{R}q_1^{(s)}, \label{eq:Pmom_S}\\
				\begin{array}{lll}
					x = 0: & \tilde{h} = 0, & T\tilde{h}_{xx} = \displaystyle\frac{12}{R}L_1q_1^{(s)}, \\
					x = 1: & \tilde{h} = 0, & T\tilde{h}_{xx} = -\displaystyle\frac{12}{R}L_2q_1^{(s)},
				\end{array} \label{eq:Pbcs_S}
			\end{gather}
			where $\mathcal{L}_s$ is the linear steady operator given by
			\begin{equation}
			\label{eq:Op_S}
				\mathcal{L}_s = T\partial^3_x + \frac{6}{5}\alpha\partial_x + \frac{36}{R}.
			\end{equation}
			
			The solution is then $\tilde{h} \equiv q_1^{(s)}h_1^{(s)}$, where $h_1^{(s)} = 1/3 + ae^{{\lambda}x} + \bar{a}e^{\bar{\lambda}x} + be^{-(\lambda + \bar{\lambda})x}$, $\lambda$ being one of the two complex roots of the characteristic equation and $a$ and $b$ are complex and real constants, respectively. Nontrivial solutions exist provided
			\begin{equation}
			\label{eq:Det_S}
				\begin{vmatrix}
					\displaystyle\frac{1}{3} & 1 & 1 & 1 \\[6pt]
					-\displaystyle\frac{12L_1}{RT} & {\lambda}^2 & {\bar{\lambda}}^2 & (\lambda + \bar{\lambda})^2 \\[6pt]
					\displaystyle\frac{1}{3} & e^{\lambda} & e^{\bar{\lambda}} & e^{-(\lambda + \bar{\lambda})} \\[6pt]
					\displaystyle\frac{12L_2}{RT} & {\lambda}^2e^{\lambda} & {\bar{\lambda}}^2e^{\bar{\lambda}} & (\lambda + \bar{\lambda})^2e^{-(\lambda + \bar{\lambda})}
				\end{vmatrix} = 0.
			\end{equation}
			
			Eigenvalue relation (\ref{eq:Det_S}) yields neutral stability curves in the $(T,R)$ plane.
			
		\subsection{\label{sec:Hopf}Hopf bifurcation}
		
			Instead of a single eigenvalue being at the origin, we assume now that if $R = R_c(T)$, $\sigma = \pm i\omega$ ($\omega > 0$), which means that a pair of complex conjugate eigenvalues crosses the imaginary axis, while others are still in the left half-plane, resulting in a bifurcation into $2\pi/\omega$ time-periodic solutions. Along the bifurcation curve, setting $\partial_t = i\omega$, we have $\tilde{h} \equiv h_1^{(o)} = AQ_{1x}(x)e^{i{\omega}t} + C.C.$ and $\tilde{q} \equiv q_1^{(o)} = -i{\omega}AQ_1(x)e^{i{\omega}t} + C.C.$ where $A$ is an arbitrary constant and $Q_1$ is an eigenfunction of the following eigenvalue problem:
			\begin{gather}
				\mathcal{L}_{\omega}Q_1 = 0, \label{eq:Pmom_O}\\
				\begin{array}{lll}
					x = 0: & Q_{1x} = 0, & \left\lbrace T\partial_x^3 + i{\omega}L_1\left(\displaystyle\frac{12}{R_c} + i\gamma\omega\right)\right\rbrace Q_1 = 0, \\
					x = 1: & Q_{1x} = 0, & \left\lbrace T\partial_x^3 - i{\omega}L_2\left(\displaystyle\frac{12}{R_c} + i\gamma\omega\right)\right\rbrace Q_1 = 0,
				\end{array} \label{eq:Pbcs_O}
			\end{gather}
			where $\mathcal{L}_{\omega}$ is the linear operator given by
			\begin{equation}
			\label{eq:Op_O}
				\mathcal{L}_{\omega} = T\partial^4_x + \frac{6}{5}\alpha\partial^2_x + \left(\frac{36}{R_c} + \frac{12}{5}i\beta\omega\right)\partial_x + \left(\frac{12}{R_c}i\omega - \gamma\omega^2\right)
			\end{equation}
			
			Hence, $Q_1(x) = \sum_{n = 1}^4 c_ne^{ik_nx}$ with the $k_n$ being the roots of the quartic characteristic equation corresponding to (\ref{eq:Pmom_O}). In order for a nontrivial solution to exist, the determinant of the linear system formed from the boundary conditions (\ref{eq:Pbcs_O}) must vanish
			$$
				det[\mathbf{v_1}, \ \mathbf{v_2}, \ \mathbf{v_3}, \ \mathbf{v_4}] = 0
			$$
			with
			$$
				\begin{array}{l}
					\mathbf{v}_j^T = \left[ik_j, -iTk_j^3 + L_1\left(\displaystyle\frac{12}{R_c}i\omega - \gamma\omega^2\right), ik_je^{ik_j}, -iTk_j^3e^{ik_j} - L_2\left(\displaystyle\frac{12}{R_c}i\omega - \gamma\omega^2\right)e^{ik_j}\right], \\
					j = 1,2,3,4.
				\end{array}
			$$
			
\begin{figure}[!t]
				\centering
				\subfloat[\label{fig:TR_Stat}]{%
					\includegraphics[width=0.485\textwidth]{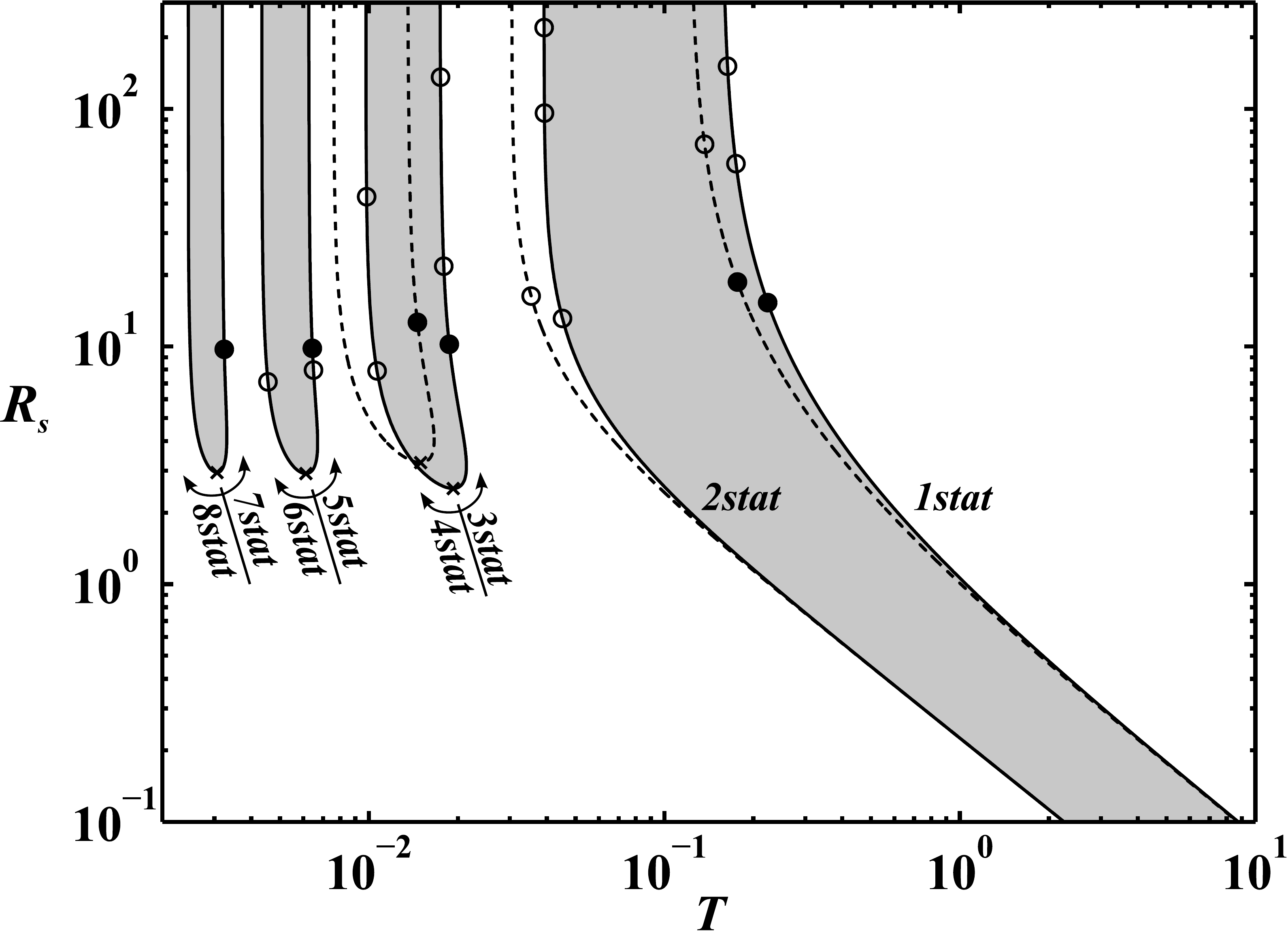}
				}\hfill
				\subfloat[\label{fig:TR_Osc}]{%
					\includegraphics[width=0.495\textwidth]{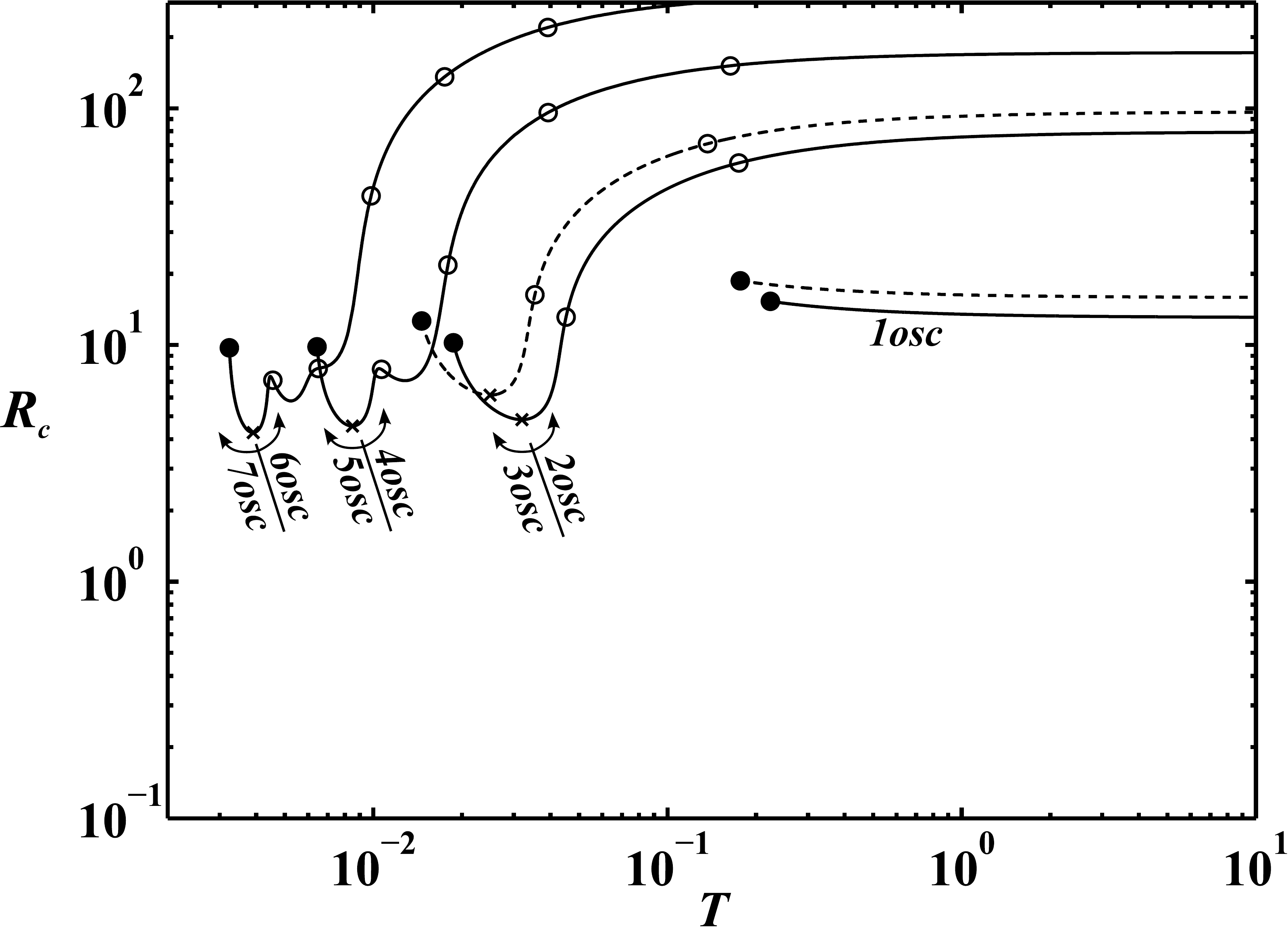}
				}
			\caption{Stability boundaries in $(T,R)$ parameter space. (a) Critical branches for monotonic (static) modes. (b) Marginal branches for oscillating modes. $n$ refers to the mode number, crosses indicate the coalescence between two successive modes while circles show the intersections of neutral and marginal curves; the Hopf frequency vanishes at points marked by dark circles. Continuous and dashed lines denote stability boundaries for the WIM and IM, respectively.}
			\label{fig:PSpace}
			\end{figure}

			This gives the curve $R = R_c(T)$ along which the fixed point $(0,0)$ loses its stability in favor of a limit cycle. Representative results on the stability boundaries in the $(T,R)$ parameter space are shown in Fig.\ \ref{fig:PSpace}, illustrating a noticeable discrepancy in the predictions of the stability boundaries between the WIM and the IM.
			
	\section{\label{sec:WeakNL}Weakly Nonlinear Analysis}
		
		Beyond criticality and far enough from particular points where more than one mode are simultaneously destabilized, the amplitude of the most unstable global mode grows to a size sufficient to excite nonlinear self-interactions which retard its exponential temporal growth until a new flow sets in.
		
		\subsection{\label{sec:SBFlow}Steady bifurcation flow}
			
			Here, the aim is to correct the linear description of the bifurcating steady flow by accounting for the self-interactions of the steady global mode which is known to oppose to its exponential growth. To do this, instead of assuming that the amplitude of the critical global mode remains arbitrarily small, we consider now that it evolves slowly with time. Hence, it appears natural to introduce a fast time $t$ and a slow time $\tau = {\delta}t$ where $\delta \ll 1$ measures the smallness of the flow amplitude near criticality. Then expanding the perturbations $(\tilde{h},\tilde{q})$ of the base state as $\tilde{h} = {\delta}q_1^{(s)}h_1^{(s)} + {\delta}^2h_2^{(s)} +\ \cdots$, $\tilde{q} = {\delta}q_1^{(s)} + {\delta}^2q_2^{(s)} +\ \cdots$, and $R = R_s(1 + {\delta}R_1)$, substituting $\partial_t$ by $\partial_t + \delta\partial_{\tau}$ in Equation (\ref{eq:Pmom}), setting $\partial_t \equiv 0$ and separating terms of like powers of $\delta$, one first obtains the linear problem (\ref{eq:Pmom_S}) at $O(\delta)$ and the following problem for $(h_2^{(s)},q_2^{(s)})$ at $O({\delta}^2)$:
			\begin{gather}
				q_{2x}^{(s)} + q_{1\tau}^{(s)}h_1^{(s)} = 0, \label{eq:Pcont_S2}\\
				\begin{aligned}
					\mathcal{L}_sh_2^{(s)} =& \frac{12}{R_s}q_2^{(s)} + \frac{12}{5}{\beta}q_{2x}^{(s)} + {\gamma}q_{1\tau}^{(s)} + \frac{12}{R_s}R_1(3h_1^{(s)} - 1)q_1^{(s)} \\ & - \left(3Th_1^{(s)}h_{1xxx}^{(s)} + \frac{36}{R_s}h_1^{(s)^2} + \frac{12}{5}{\alpha}h_{1x}^{(s)}\right)q_1^{(s)^2}
				\end{aligned} \label{eq:Pmom_S2}
			\end{gather}
			subject to
			\begin{equation}
			\label{eq:Pbcs_S2}
				\begin{array}{lll}
				x = 0: & h_2^{(s)} = 0, & Th_{2xx}^{(s)} - L_1\left\lbrace \displaystyle\frac{12}{R_s}(q_2^{(s)} - R_1q_1^{(s)}) + {\gamma}q_{1\tau}^{(s)} \right\rbrace = 0, \\
				x = 1: & h_2^{(s)} = 0, & Th_{2xx}^{(s)} + L_2\left\lbrace \displaystyle\frac{12}{R_s}(q_2^{(s)} - R_1q_1^{(s)}) + {\gamma}q_{1\tau}^{(s)} \right\rbrace = 0.
			\end{array}
			\end{equation}
			
			Here, the amplitude of $q_1^{(s)}$ is assumed to depend on the slow time $\tau$. Thus, solving (\ref{eq:Pcont_S2}) for $q_2^{(s)}$ gives
			\begin{equation}
			\label{eq:Pcont_Sq2}
				q_2^{(s)} = -q_{1\tau}^{(s)}\int h_1^{(s)}(x) \mathrm{d}x.
			\end{equation}
			
			The integration constant is dropped since it can be incorporated in the definition of $q_1^{(s)}(\tau)$. Then
			substituting the above expression for $q_2^{(s)}$ in the right hand side of (\ref{eq:Pmom_S2}) yields
			\begin{equation}
			\label{eq:Pmom_S2s}
				\mathcal{L}_sh_2^{(s)} = S(x),
			\end{equation}
			where
			\begin{equation}
			\label{eq:Psource_S2}
				\begin{aligned}
					S(x) = &\left(\gamma - \frac{12}{R_s}\int h_1^{(s)} \mathrm{d}x - \frac{12}{5}\beta h_1^{(s)}\right)q_{1\tau}^{(s)} + \frac{12}{R_s}(3h_1^{(s)} - 1)R_1q_1^{(s)} \\ & - \left(3Th_1^{(s)}h_{1xxx}^{(s)} + \frac{36}{R_s}h_1^{(s)^2} + \frac{12}{5}{\alpha}h_{1x}^{(s)}\right)q_1^{(s)^2}.
				\end{aligned}
			\end{equation}
			
			The kernel of the operator $\mathcal{L}_s$ is not empty, therefore, for a solution to (\ref{eq:Pmom_S2s}) to exist, the compatibility condition $S(x) \in Image(\mathcal{L}_s)$ is required. Practically, this condition is expressed by invoking the solvability theorem for boundary value problems; i.e., the orthogonality (in the sense of some inner product to be defined) of $S(x)$ and the spanning function $h^{(a)}$ of the kernel of the adjoint operator $\mathcal{L}^{(a)}_s$ associated to $\mathcal{L}_s$, namely,
			\begin{equation}
			\label{eq:In_Prod1}
				\langle h^{(a)},S\rangle = 0,
			\end{equation}
			where the inner product is defined as (see Appendix \ref{app:AdjProb1} for derivation)
			\begin{equation}
			\label{eq:In_Prod2}
				\langle h^{(a)},S \rangle = \int_0^1 h^{(a)}(x)S(x)\mathrm{d}x + S(0)h^{(a)}(0) + S(1)h^{(a)}(1),
			\end{equation}
			with
			\begin{equation}
			\label{eq:In_Prod3}
				S(0) = Th_{2xx}^{(s)}(0), \quad\quad S(1) = -Th_{2xx}^{(s)}(1).
			\end{equation}
			
			Equation (\ref{eq:In_Prod1}) along with Equations (\ref{eq:In_Prod2}), (\ref{eq:In_Prod3}), and (\ref{eq:Pbcs_S2}) yields
			\begin{equation}
			\label{eq:AdjSP_S2}
				\begin{aligned}
					& \int_0^1 S(x)h^{(a)} \mathrm{d}x + \left(-\frac{12}{R_s}R_1q_1^{(s)} + {\gamma}q_{1\tau}^{(s)}\right)(L_1h^{(a)}(0) + L_2h^{(a)}(1)) \\ & + \frac{12}{R_s}\left(L_1h^{(a)}q_2^{(s)}\Big{|}_0 + L_2h^{(a)}q_2^{(s)}\Big{|}_1\right) = 0.
				\end{aligned}
			\end{equation}
			
			Combining this equation with (\ref{eq:Psource_S2}) yields the following amplitude equation:
			\begin{equation}
			\label{eq:Amp_S1}
				\frac{{\partial}q_1^{(s)}}{\partial\tau} = \alpha_sR_1q_1^{(s)} - \beta_s{q_1^{(s)}}^2
			\end{equation}
			which reads, after the change ${\delta}q_1^{(s)} \rightarrow q_1^{(s)}$,
			\begin{equation}
			\label{eq:Amp_S2}
				\frac{{\partial}q_1^{(s)}}{\partial{t}} = \alpha_s\frac{R-R_s}{R_s}q_1^{(s)} - \beta_s{q_1^{(s)}}^2.
			\end{equation}
			
			This equation characterizes a transcritical bifurcation with stability exchange. The coefficient $\alpha_s(R/R_s - 1)$ is the exponential growth rate of the critical monotonic global mode while $\beta_s$ measures its quadratic self-interaction which is known to moderate its exponential growth during the linear stage.
			
		\subsection{\label{sec:OBFlow}Oscillatory bifurcation flow}
			
			Here, we apply the procedure developed above to the Hopf bifurcation with frequency $\omega$ corresponding to $R = R_c(T)$. It will turn out in the following that slightly past the threshold, the relevant slow time is $\xi = {\delta}^2t$ and the corresponding expansion of $R$ is $R = R_c(1 + {\delta}^2R_2 +\ \cdots)$. The slow time $\tau = {\delta}t$ and the $O(\delta)$ term in the expansion of $R$ would cancel due to the absence, at $O({\delta}^2)$, of resonant terms. Then substituting for $\partial_t$ the operator $\partial_t + {\delta}^2\partial_\xi$, expanding $(\tilde{h},\tilde{q}) = {\delta}(h_1^{(o)},q_1^{(o)}) + {\delta}^2(h_2^{(o)},q_2^{(o)}) + {\delta}^3(h_3^{(o)},q_3^{(o)}) +\ \cdots$, and identifying successive powers of $\delta$ in Equation (\ref{eq:Imom}) yields a sequence of three linear problems. The former is the marginal problem for $(h_1^{(o)},q_1^{(o)})$ whose solution is already calculated. At $O(\delta^2)$ we get
			\begin{gather}
				h_{2t}^{(o)} + q_{2x}^{(o)} = 0, \label{eq:Pcont_O2}\\
				\begin{aligned}
					& Th_{2xxx}^{(o)} + \frac{6}{5}{\alpha}h_{2x}^{(o)} + \frac{12}{R_c}(3h_2^{(o)} - q_2^{(o)}) - \frac{12}{5}{\beta}q_{2x}^{(o)} - {\gamma}q_{2t}^{(o)} \\ &= S_{20}(x)|A(\xi)|^2 + S_{22}(x)A^2(\xi)e^{2i{\omega}t} + C.C.,
				\end{aligned} \label{eq:Pmom_O2}
			\end{gather}
			with boundary conditions
			\begin{equation}
			\label{eq:Pbcs_O2}
				\begin{array}{lll}
					x = 0: & h_2^{(o)} = 0, & Th_{2xx}^{(o)} - L_1\left(\displaystyle\frac{12}{R_c}q_2^{(o)} + {\gamma}q_{2t}^{(o)}\right) = 0, \\
					x = 1: & h_2^{(o)} = 0, & Th_{2xx}^{(o)} + L_2\left(\displaystyle\frac{12}{R_c}q_2^{(o)} + {\gamma}q_{2t}^{(o)}\right) = 0.
				\end{array}
			\end{equation}
			
			The nonhomogeneous term in (\ref{eq:Pmom_O2}) is related to the eigensolution ($h_1^{(o)},q_1^{(o)}$) obtained in Section \hyperref[sec:Hopf]{\ref*{sec:LinStab}.\ref*{sec:Hopf}}, namely,
			\begin{gather}
				S_{20}(x) = -3T\bar{Q}_{1x}Q_{1xxxx} - \frac{12}{5}i\alpha\omega\bar{Q}_1Q_{1xx} - \frac{36}{R_c}|Q_{1x}|^2 + 2{\omega}^2\left(\gamma + \frac{6}{5}\beta\right)\bar{Q}_1Q_{1x}, \label{eq:Psource1_O2}\\
				S_{22}(x) = -3TQ_{1x}Q_{1xxxx} + \frac{12}{5}i\alpha{\omega}Q_1Q_{1xx} - \frac{36}{R_c}Q_{1x}^2 + 2{\omega}^2\left(\gamma - \frac{6}{5}\beta\right)Q_1Q_{1x}. \label{eq:Psource2_O2}
			\end{gather}
			
			We note that (\ref{eq:Pmom_O2}) does not contain time-resonant terms, therefore it can be solved without any subsidiary condition. The solution of (\ref{eq:Pmom_O2}) is decomposed into steady and oscillatory parts as
			\begin{gather}
				q_2^{(o)} = |A(\xi)|^2Q_{20} - 2i{\omega}A^2(\xi)Q_{22}(x)e^{2i{\omega}t} + C.C., \label{eq:50}\\
				h_2^{(o)} = |A(\xi)|^2H_{20}(x) + A^2(\xi)Q_{22x}e^{2i{\omega}t} + C.C. \label{eq:51}
			\end{gather}
			
			Solving (\ref{eq:Pcont_O2}) and (\ref{eq:Pmom_O2}) for $H_{20}(x)$ and $Q_{20}$, one obtains
			\begin{equation}
			\label{eq:sol_H20}
				H_{20}(x) = \frac{Q_{20}}{3} + c_{21}^*e^{{\lambda}x} + \bar{c}_{21}^*e^{\bar{\lambda}x} + c_{22}^*e^{-({\lambda} + \bar{\lambda})x}
			\end{equation}
			with
			$$
				c_{2j}^*(x) = \tilde{c}_{2j}(x) + c_{2j}, \qquad j = 1,2,
			$$
			where
			\begin{equation}
			\label{eq:Const_H20}
				\begin{array}{lll}
					\tilde{c}_{21}(x) = \displaystyle\frac{1}{T}\frac{\displaystyle\int_0^x(S_{20} + \bar{S}_{20})e^{-{\lambda}x} \mathrm{d}x}{2{\lambda}^2 - \bar{\lambda}({\lambda} + \bar{\lambda})} & \text{and} & \tilde{c}_{22}(x) = \displaystyle\frac{1}{T}\frac{\displaystyle\int_0^x(S_{20} + \bar{S}_{20})e^{({\lambda} + \bar{\lambda})x} \mathrm{d}x}{2({\lambda} + \bar{\lambda})^2 - \lambda\bar{\lambda}}
				\end{array}
			\end{equation}
			and $Q_{20}$, $c_{21}$, $\bar{c}_{21}$, and $c_{22}$ are yet to be determined constants by forcing the solution to satisfy the boundary conditions. We are led to a fourth order linear nonhomogeneous algebraic system of the form
			\begin{equation}
			\label{eq:Const_Q20}
				\mathbf{B}[Q_{20} \ c_{21} \ \bar{c}_{21} \ c_{22}]^T = \mathbf{b}
			\end{equation}
			with
			$$
				\mathbf{B} =
				\begin{bmatrix}
					\displaystyle\frac{1}{3} & 1 & 1 & 1 \\[6pt]
					-\displaystyle\frac{12L_1}{TR_c} & \lambda^2 & \bar{\lambda}^2 & (\lambda + \bar{\lambda})^2 \\[6pt]
					\displaystyle\frac{1}{3} & e^{\lambda} & e^{\bar{\lambda}} & e^{(\lambda + \bar{\lambda})} \\[6pt]
					\displaystyle\frac{12L_2}{TR_c} & \lambda^2e^{\lambda} & \bar{\lambda}^2e^{\bar{\lambda}} & (\lambda + \bar{\lambda})^2e^{(\lambda + \bar{\lambda})}
				\end{bmatrix},
			$$
			$$
				\mathbf{b} = 
				-\begin{bmatrix}
					0 \\ \\
					0 \\ \\
					\tilde{c}_{21}(1)e^{\lambda} + \bar{\tilde{c}}_{21}(1)e^{\bar{\lambda}} + \tilde{c}_{22}(1)e^{-(\lambda+\bar{\lambda})} \\ \\
					\tilde{c}_{21}(1)\lambda^2e^{\lambda} + \bar{\tilde{c}}_{21}(1)\bar{\lambda}^2e^{\bar{\lambda}} + \tilde{c}_{22}(1)(\lambda + \bar{\lambda})^2e^{-(\lambda+\bar{\lambda})}
				\end{bmatrix}.
			$$
			
			Now ignoring the steady term in (\ref{eq:Pcont_O2})-(\ref{eq:Pbcs_O2}), one obtains for $Q_{22}$,
			\begin{equation}
			\label{eq:Eqn_Q22}
				\mathcal{L}_{2\omega}Q_{22} = S_{22}(x)
			\end{equation}
			subject to
			\begin{equation}
			\label{eq:BCs_Q22}
				\begin{array}{c}
					Q_{22x}(0) = Q_{22x}(1) = 0, \\
					\left[TQ_{22xxx} + L_1\left(\displaystyle\frac{24}{R_c}i\omega - 4\gamma\omega^2\right)Q_{22}\right]_0 = 0, \\[10pt]
					\left[TQ_{22xxx} - L_2\left(\displaystyle\frac{24}{R_c}i\omega - 4\gamma\omega^2\right)Q_{22}\right]_1 = 0.
				\end{array}
			\end{equation}
			
			In Equation (\ref{eq:Eqn_Q22}), $\mathcal{L}_{2\omega}$ denotes the operator obtained from $\mathcal{L}_{\omega}$ in Equation (\ref{eq:Op_O}) once $\omega$ is replaced by $2\omega$. Expressing $Q_{22}(x)$ in the form
			\begin{equation}
			\label{eq:sol_Q22}
				\begin{array}{lll}
					Q_{22}(x) = \sum_{n=1}^4[a_{2n}^*(x) + a_{2n}]e^{ik_nx} & \text{with} & a_{2n}^*(x) = \displaystyle\frac{1}{T}\frac{cof_{4n}(\mathbf{W})}{det(\mathbf{W})}\int_0^x S_{22}(x)e^{-ik_nx} \mathrm{d}x,
				\end{array}
			\end{equation}
			where $cof_{4n}(\mathbf{W})$ refers to the cofactor of the matrix $\mathbf{W}$ below resulting from deletion of the $4$th row and $n$th column and $a_{2n}$ are yet to be determined constants by using the above boundary conditions,
			$$
				\mathbf{W} = 
				\begin{bmatrix}
					1 & 1 & 1 & 1 \\
					ik_{21} & ik_{22} & ik_{23} & ik_{24} \\[2pt]
					-k_{21}^2 & -k_{22}^2 & -k_{23}^2 & -k_{24}^2 \\[2pt]
					-ik_{21}^3 & -ik_{22}^3 & -ik_{23}^3 & -ik_{24}^3
				\end{bmatrix}.
			$$
			
			So we get the following system:
			\begin{equation}
			\label{eq:Const_Q22}
				[\mathbf{w}_1, \ \mathbf{w}_2, \ \mathbf{w}_3, \ \mathbf{w}_4]
				\begin{bmatrix}
					a_{21} \\ a_{22} \\ a_{23} \\ a_{24}
				\end{bmatrix} =
				\begin{bmatrix}
					0 \\ 0 \\ \displaystyle -\sum ik_{2j}a_{2j}^*(1)e^{ik_{2j}} \\[3pt] \displaystyle T\sum ik_{2j}^3a_{2j}^*(1)e^{ik_{2j}} + L_2\left(\frac{24}{R_c}i\omega - 4\gamma\omega^2\right)\sum a_{2j}^*(1)e^{ik_{2j}}
				\end{bmatrix},
			\end{equation}
			where
			$$
				\mathbf{w}_j^T = \left[ik_{2j}, -iTk_{2j}^3 + L_1\left(\frac{24}{R_c}i\omega - 4\gamma\omega^2\right), ik_{2j}e^{ik_{2j}}, -iTk_{2j}^3e^{ik_{2j}} - L_2\left(\frac{24}{R_c}i\omega - 4\gamma\omega^2\right)e^{ik_{2j}}\right].
			$$
			
			The fast time dependence of quadratic self-interactions of marginal oscillatory global modes never has the frequency $\pm\omega$ which would lead to resonance. Accordingly, no quadratic terms occur in the amplitude equations. Therefore, the evolution equation for the amplitude $A(\xi)$ has to be formed at the next order which reads
			\begin{gather}
				h_{3t}^{(o)} + q_{3x}^{(o)} = -h_{1\xi}^{(o)}, \label{eq:Pcont_O3}\\
				Th_{3xxx}^{(o)} + \frac{6}{5}{\alpha}h_{3x}^{(o)} + \frac{12}{R_c}(3h_3^{(o)} - q_3^{(o)}) - \frac{12}{5}{\beta}q_{3x}^{(o)} - {\gamma}q_{3t}^{(o)} = S_3(x)e^{i{\omega}t} + C.C. + N.R.T., \label{eq:Pmom_O3}
			\end{gather}
			where $N.R.T.$ stands for non-resonant terms. They do not enter in the following calculation, so we refrain from giving them here because they are too cumbersome. The resonant term has the form
			\begin{equation}
			\label{eq:Psource_O3}
				S_{3}(x) = S_{31}(x)A_\xi + S_{32}(x)R_2A + S_{33}(x)A|A|^2,
			\end{equation}
			where $S_{31}$, $S_{32}$, and $S_{33}$ are expressed in Appendix \ref{app:Source}.
			
			Solving the system (\ref{eq:Pcont_O3}) and (\ref{eq:Pmom_O3}) by considering only the resonant source term, we can eliminate $h_3^{(o)}$ from (\ref{eq:Pcont_O3}) by setting $\partial_t = i\omega$ and $q_3^{(o)} = -i{\omega}Q_3(x)e^{i{\omega}t}$, Equation (\ref{eq:Pmom_O3}) then reduces to
			\begin{equation}
			\label{eq:Pmom_O3s}
				\mathcal{L}_{\omega}Q_3 = \tilde{S}_3(x)
			\end{equation}
			subject to
			\begin{equation}
			\label{eq:Pbcs_O3}
				\begin{array}{l}
					Q_{3x}\Big{|}_0 = 0, \qquad Q_{3x}\Big{|}_1 = 0, \\
					TQ_{3xxx}\Big{|}_0 + L_1\left(\displaystyle\frac{12}{R_c}i\omega - \gamma\omega^2\right)Q_3\Big{|}_0 = B_0, \\
					TQ_{3xxx}\Big{|}_1 - L_2\left(\displaystyle\frac{12}{R_c}i\omega - \gamma\omega^2\right)Q_3\Big{|}_1 = -B_1,
				\end{array}
			\end{equation}
			where
			$$
				\tilde{S}_3(x) = S_3(x) - \left[\frac{12}{5}{\beta}Q_{1x} + \left(\frac{12}{R_c} + i\gamma\omega\right)Q_1\right]A_\xi,
			$$
			$$
				\begin{array}{l}
					B_0 = \displaystyle L_1\left[\frac{12}{R_c}i{\omega}R_2A - \left(\frac{12}{R_c} + 2i\gamma\omega\right)A_{\xi}\right]Q_1\Big{|}_0, \\[10pt]
					B_1 = \displaystyle L_2\left[\frac{12}{R_c}i{\omega}R_2A - \left(\frac{12}{R_c} + 2i\gamma\omega\right)A_{\xi}\right]Q_1\Big{|}_1.
				\end{array}
			$$
			
			The evolution equation for $A(\xi)$ will be found once again by invoking the Fredholm solvability condition (see Appendix \ref{app:AdjProb2}). One obtains the so-called Landau equation
			\begin{equation}
				\frac{{\partial}A}{\partial\xi} = \alpha_cR_2A - \beta_c|A|^2A
			\end{equation}
			which can be rewritten by the change ${\delta}A \rightarrow A$ as
			\begin{equation}
				\frac{{\partial}A}{{\partial}t} = \alpha_c\frac{R - R_c}{R_c}A - \beta_c|A|^2A.
			\end{equation}
			Here the coefficient $\beta_c$ measures the cubic self-interaction of the critical oscillatory global mode. Its sign indicates the subcritical or supercritical nature of the bifurcation.
			
	\section{\label{sec:Num}Numerics}
		
\begin{figure}[!h]
			\centering
			\subfloat[\label{fig:h1_Stat}]{%
				\includegraphics[width=0.502\textwidth]{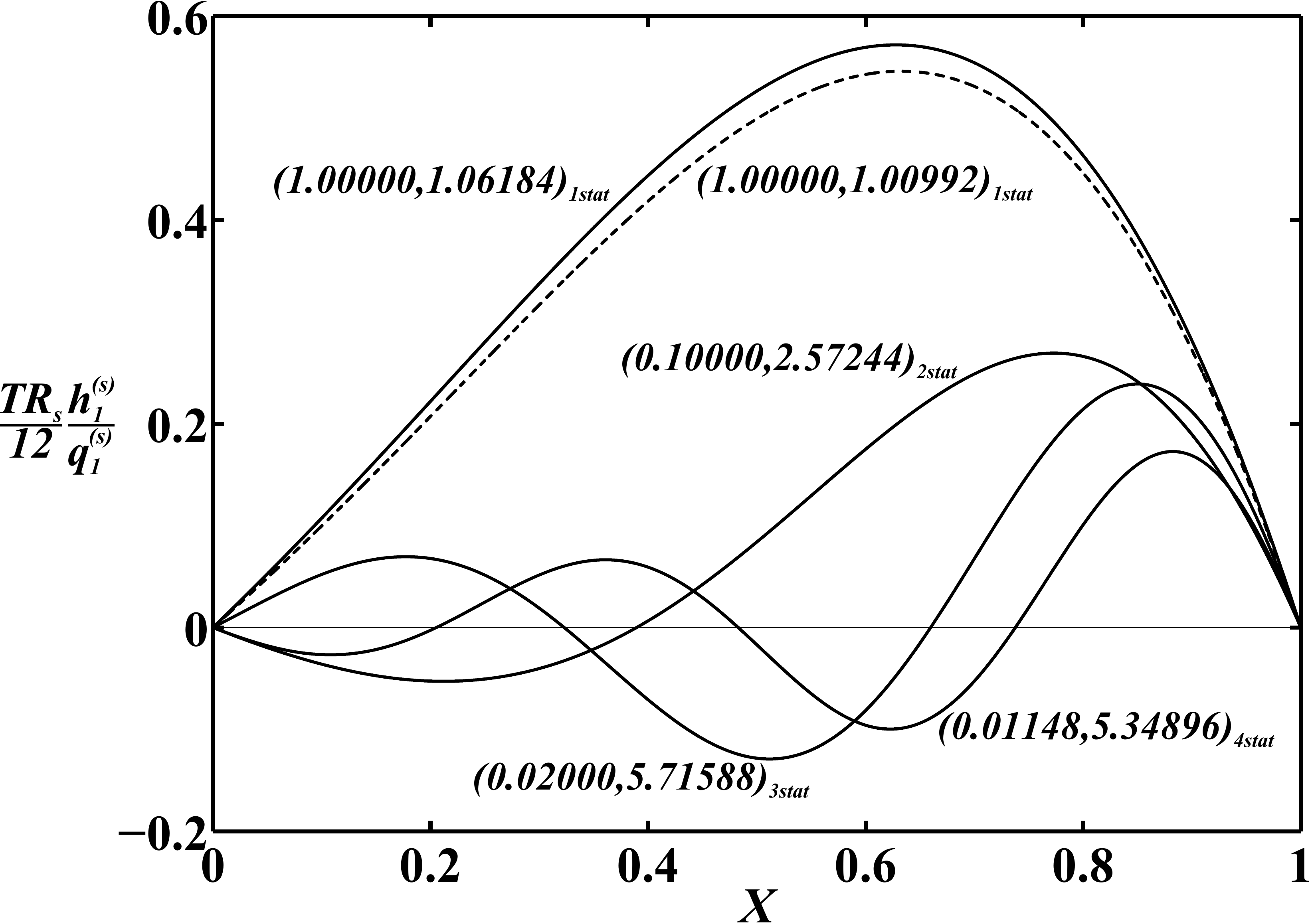}
			}\hfill
			\subfloat[\label{fig:Qx_Osc}]{%
				\includegraphics[width=0.48\textwidth]{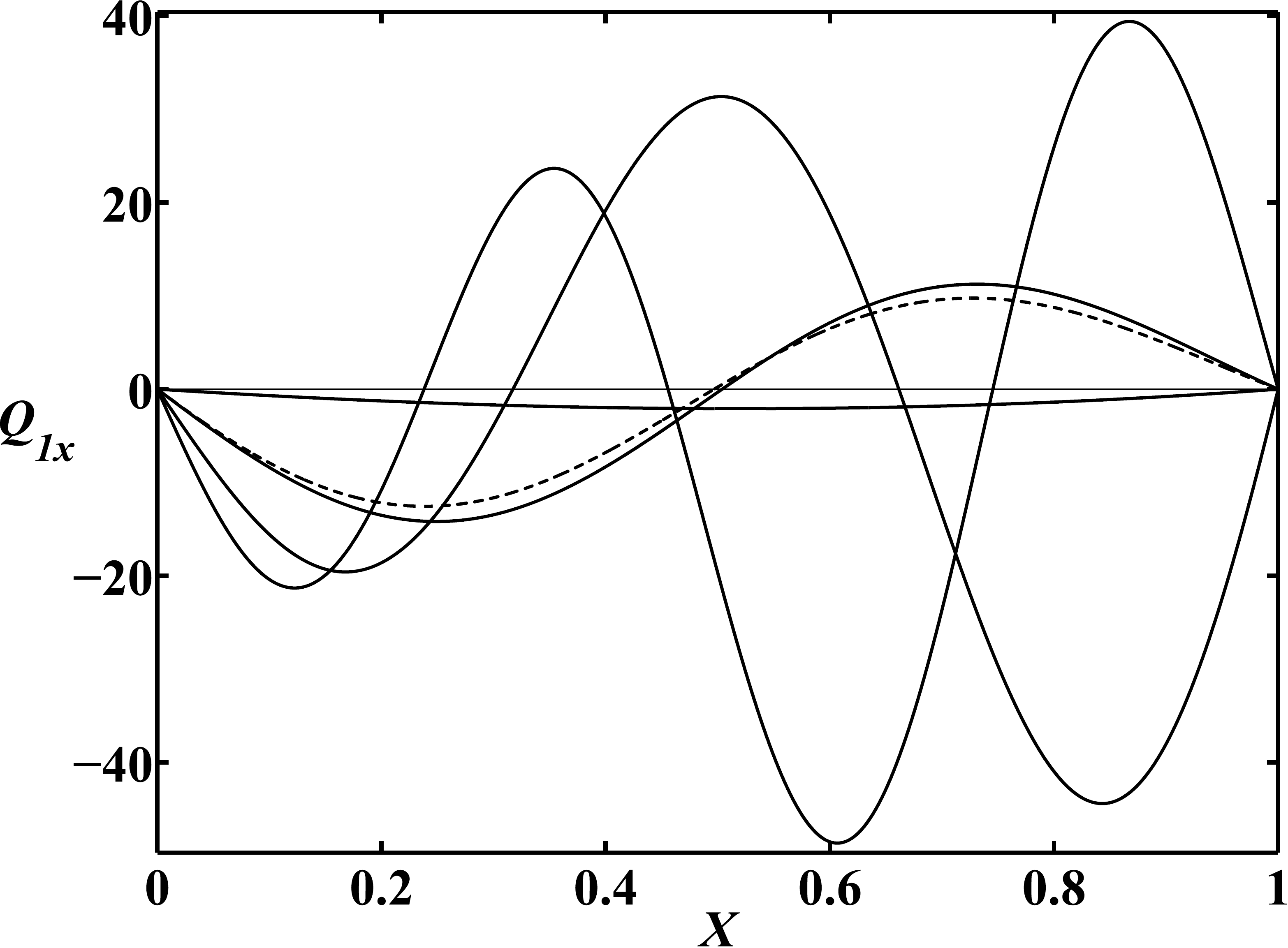}
			}
			\caption{(a) Static and (b) oscillating eigenmode shapes for the WIM (continuous lines). The mode number refers to the number of extrema of the mode shape and characterizes each critical branch. Dashed lines correspond to the mode shape of the IM with the same membrane tension as the corresponding mode shape for the WIM.}
			\label{fig:HQ}
		\end{figure}

		Following \citeA{Stewart09}, we assume throughout this section that $L_2$ is much greater than $L_1$ and set $L_2 = 10$ and $L_1 = 1$. Besides the fact that this configuration promotes oscillatory instability, as shown by these authors, it also allows for comparisons to be made. The curves drawn in Fig.\ \ref{fig:TR_Stat} compare the neutral stability boundaries (corresponding to a real eigenvalue crossing the imaginary axis) in the $(T,R)$ parameter space obtained from the WIM (continuous lines) and the IM (dashed lines). They exhibit a noticeable discrepancy in the prediction of the monotonic instability threshold beyond some level of inertia. The departure between the two models becomes asymptotically small as $R \rightarrow 0$. Shaded regions indicate where the base flow is unstable against monotonic (static) modes and where a bifurcating steady flowsets in via a transcritical bifurcation. For small to moderate values of $T$, the onset of instability appears by decreasing or increasing $R$ depending on the shape (mode number) of the unstable mode. Each unstable region is bounded by two neutral stability branches termed ``$n$stat'' and ``$(n+1)$stat'' ($n = 1, 3, 5, 7, ...$) indicating the mode number of the corresponding unstable mode. At the intersection points (marked by crosses), the modes $n$ and $n + 1$ coalesce. The mode number $n$, which naturally decreases with increasing $T$, refers to the number of extrema in the spatial structures of the critical modes as shown in Fig.\ \ref{fig:h1_Stat}. It is to be noted that monotonic instability occurs primarily for relatively small values of $T$, except when $R \rightarrow 0$ where higher values of $T$ are required to promote instability. A bifurcating steady flow may then take place in narrow bands provided the Reynolds number exceeds critical values which are approximately $2$ and $3$ for the modes $3$, ..., $8$. Outside these bands, the pressure exerted by the tensioned membrane overcomes the destabilizing effect of inertia. However, for high enough Reynolds numbers inertia dominates the pressure effects and the instability changes character to become oscillatory and remains so for all values of $T$ and $R$.
		
		\begin{figure}[!h]
			\centering
			\includegraphics[width=0.6\textwidth]{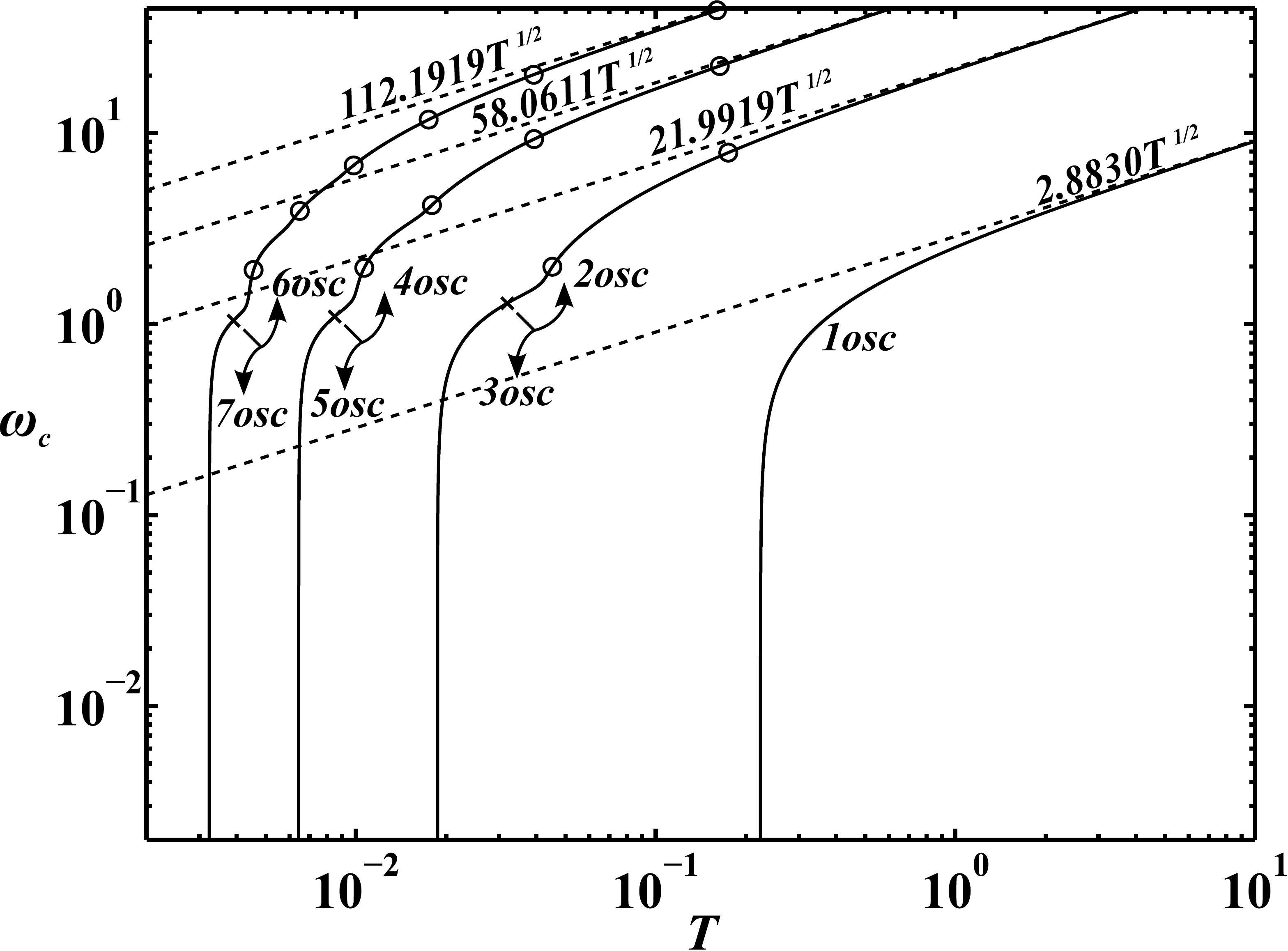}
		\caption{Variation of Hopf frequency with membrane tension of the first seven oscillating eigenmodes showing a power law behavior for large values of $T$.}
		\label{fig:Tw_Osc}
		\end{figure}
		
		Besides the stationary bifurcating flow, a time-periodic flow may also bifurcate from the base solution via a Hopf bifurcation. In the $(T,R)$ plane, this occurs along marginal curves termed ``$n$osc'' (see Fig.\ \ref{fig:TR_Osc}), $n$ being the oscillatory mode number which also refers to the number of extrema in the shape of the critical mode (see Fig.\ \ref{fig:Qx_Osc}). Fig.\ \ref{fig:TR_Osc} shows a strong dependence of the marginal Reynolds number $R_c$ on the membrane tension as $T$ falls to small values whereas it remains almost constant as $T$ is increased beyond some limit of order unity, regardless of the mode number. The base flow loses stability, in favor of a $2\pi/\omega$ time-periodic flow, when the critical branches are crossed by increasing $R$ with $T$ being fixed. Crosses indicate points where two critical modes of mode numbers ($2$,$3$), ($4$,$5$), and ($6$,$7$) coalesce. For example, modes ($2$,$3$) coalesce at $(T,R) = (0.03198,4.8388)$ or $(0.02493,6.1429)$ when using the WIM or IM, respectively. Where oscillatory and static critical curves intersect each other, monotonic and oscillatory modes interact and both bifurcate, which may result in complex dynamics and transitions. These co-dimension $2$ points are denoted by open and dark circles in Fig.\ \ref{fig:PSpace}. Points represented by dark circles mark the beginning of the marginal branches where the Hopf frequency vanishes (double-zero eigenvalue). Their proximity complicates the overall bifurcation structure which can be described by the Takens-Bogdanov normal form \cite{Guckenheimer83}. Table \ref{tab:TB_Points} compares, for the sake of illustration, the coordinates of the Takens-Bogdanov points in the $(T,R)$ plane obtained from the WIM and IM. Highly elaborate local and weakly nonlinear analyses by \citeA{Xu13} reveal how intricate the dynamics is around the particular Takens-Bogdanov point ($R \rightarrow \infty$, $T = T_{20}$), where oscillatory mode $2$ emerges through an interaction between two static eigenmodes when $L_2 = O(1)$, before ultimately growing to large amplitude. The same organizing center was also investigated by \citeA{Xu14} but in the case of a longer downstream channel. This additional degree of freedom enables the system to sustain oscillations arising from resonance between mode $2$ and mode $1$ unstable oscillations.
		
		\begin{table}[!h]
			\caption{Takens-Bogdanov points tabulated for successive modes for both the WIM and IM.}
			\begin{center}
				\begin{tabular}{lcccc}
					& & & & \\[-2.5em]
					\hline\hline\\[-1.0em]
					\multicolumn{5}{c}{Takens-Bogdanov points $(T,R)$ for the WIM and IM} \\
					\hline\\[-1.0em]
					Mode Number ($n$) & $1$ & $3$ & $5$ & $7$ \\
					WIM & $(0.2237,15.30)$ & $(0.0187,10.23)$ & $(0.0064,9.83)$ & $(0.0032,9.73)$ \\
					IM & $(0.1766,18.68)$ & $(0.0146,12.63)$ & $(0.0050,12.18)$ & $(0.0025,12.05)$ \\
					\hline\hline
				\end{tabular}
			\end{center}
			\label{tab:TB_Points}
		\end{table}
		
		\begin{figure}[!b]
			\centering
			\subfloat[\label{fig:Comp_TR}]{%
				\includegraphics[width=0.49\textwidth]{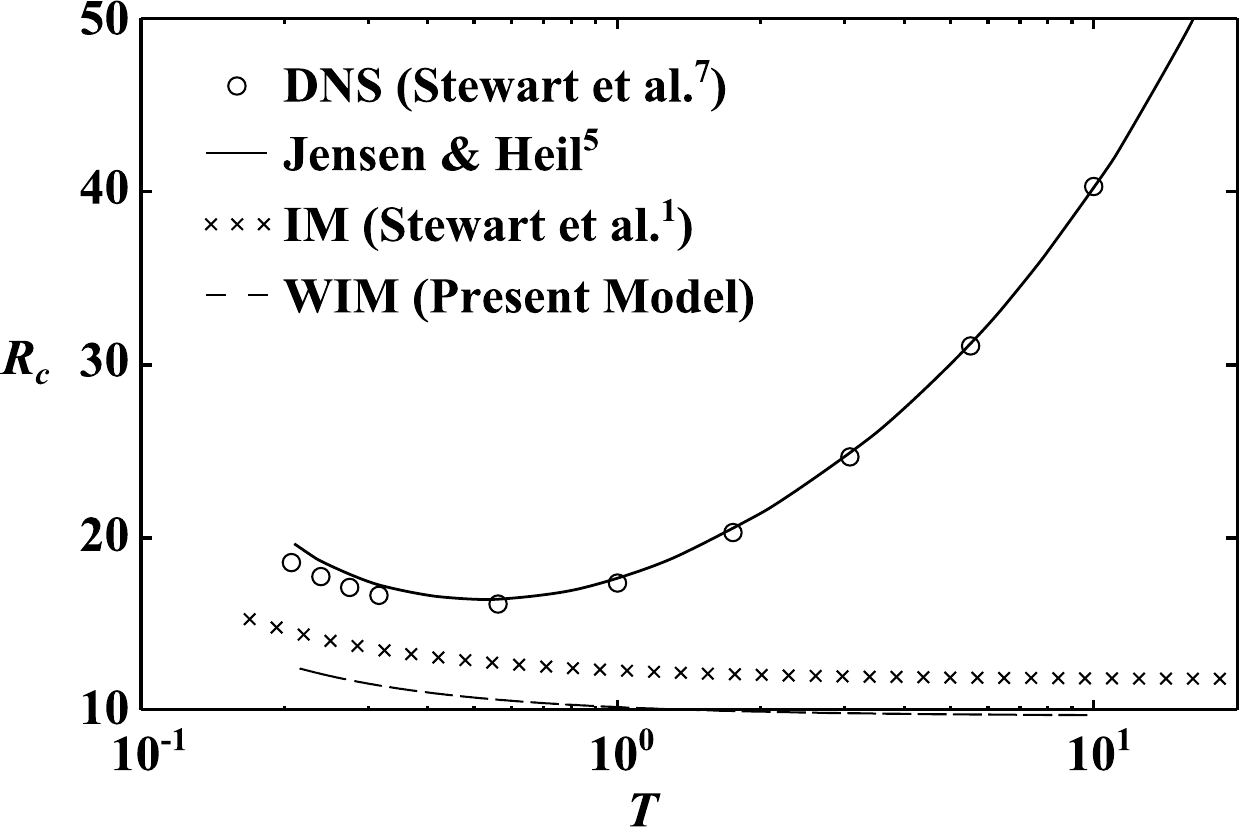}
			}\hfill
			\subfloat[\label{fig:Comp_Tw}]{%
				\raisebox{-1.24\height}{\includegraphics[width=0.49\textwidth]{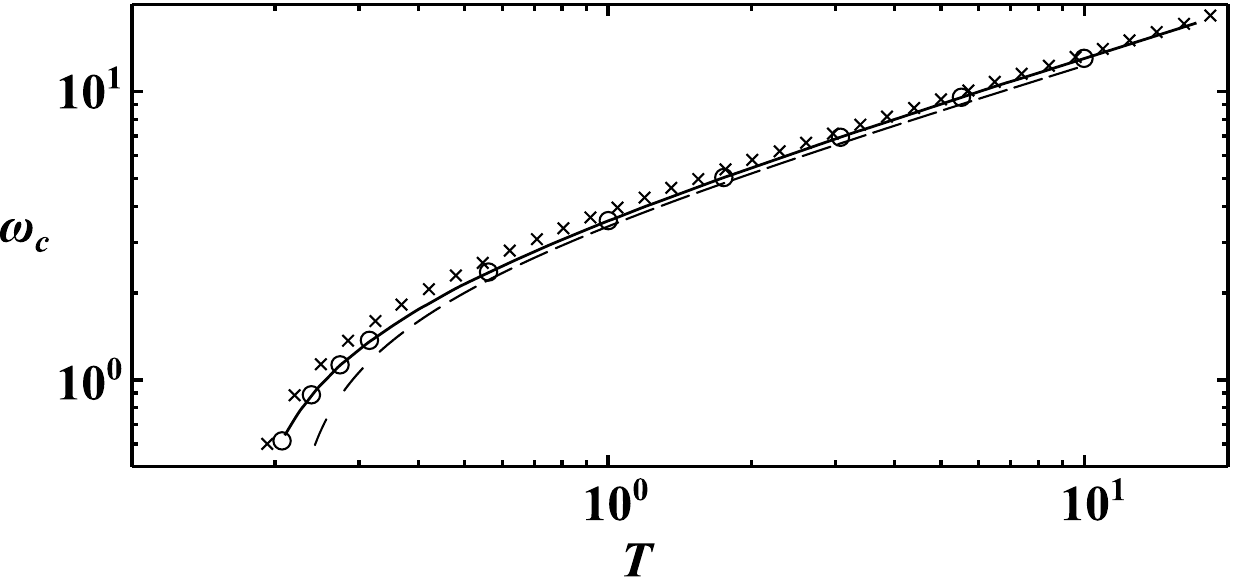}}
			}
		\caption{(a) Neutral stability boundary in the $(T,R)$ plane for the oscillatory mode $1$ for $L_1 = 0.5$ and $L_2 = 3$ and (b) corresponding frequencies versus $T$. The open circles show the results from direct numerical simulation of \protect\citeA{Stewart10} and the solid line shows the predictions of the model reported in \protect\citeA{Jensen03}. The crosses indicate the predictions from the integral model of \protect\citeA{Stewart09} and the dashed line represents the weighted integral model of the present paper.}
		\label{fig:Comp}
		\end{figure}
		
		Marginal curves for oscillatory modes $1$, $2$, $4$, and $6$ asymptote to $R = 13.07$, $79.61$, $172.17$, and $308.76$, respectively, as $T \rightarrow \infty$. These values are found, from an asymptotic analysis similar to that employed in \citeA{Stewart09}, to be $\gamma^{-1}$ times the corresponding values obtained by the IM. The variations of marginal mode frequencies $\omega$ with $T$ are plotted in Fig.\ \ref{fig:Tw_Osc}. Rapid variations which are first observed for small $T$ are followed by asymptotic behavior of the form $a_n(\gamma^{-1}T)^{1/2}$, where $a_n$ indicates the sequence of eigenvalues of the leading order problem associated to (\ref{eq:Pmom_O}) and (\ref{eq:Pbcs_O}) for sufficiently high values of $T$ (for the first two modes, one has $a_1 = 3.1807$ and $a_2 = 24.1251$ for example), namely,
		$$
			\left(\partial^4_x - a^2_n\right)Q_1 = 0 \qquad \text{subject to} \qquad Q_{1x}\Big{|}_{0,1} =
			\left(\partial^3_x - L_1a^2_n\right)Q_1\Big{|}_0 = \left(\partial^3_x + L_2a^2_n\right)Q_1\Big{|}_1 = 0.
		$$
		
		In Fig.\ \ref{fig:Comp}, we compare the predictions of the critical Reynolds number and the corresponding critical frequency of the WIM and the IM with those given by direct numerical simulation in the same conditions explored by \citeA{Stewart10} (with the present scaling, the corresponding rigid channel lengths are $L_1 = 0.5$ and $L_2 = 3$). It is clear that the model of \citeA{Jensen03} is the most appropriate one to predict the instability threshold for the mode $1$ oscillation at sufficiently high membrane tension. At lower tension, less than about $100$, the IM and WIM are rather better candidates. The WIM is expected, by virtue of its derivation, to perform, if not better, at least as well as the IM does. However, even if good and equivalent estimates of the critical frequency are predicted by the two one-dimensional models, we note that the IM predicts the critical Reynolds number more accurately than the WIM. This apparent loss of accuracy of the WIM is expected to be caused by the fact that the viscous term $u_{xx}$ is incorporated in the DNS but is not taken into account in the models. Incorporating this second order term and using a suitable velocity profile for coherence would be a possible way to better appreciate the relative accuracies of the two integral models.
		
		Typical monotonic and oscillatory eigenfunctions corresponding to modes $1$ through $4$ are represented in Fig.\ \ref{fig:HQ}. As emphasized by \citeA{Heil03}, we observe that the steady membrane motions have larger amplitudes towards the downstream end of the membrane while time-periodic flows yield larger flow rates at the upstream end.
		
\begin{figure}[!h]
			\centering
			\subfloat[\label{fig:a_Stat}]{%
				\includegraphics[width=0.44\textwidth]{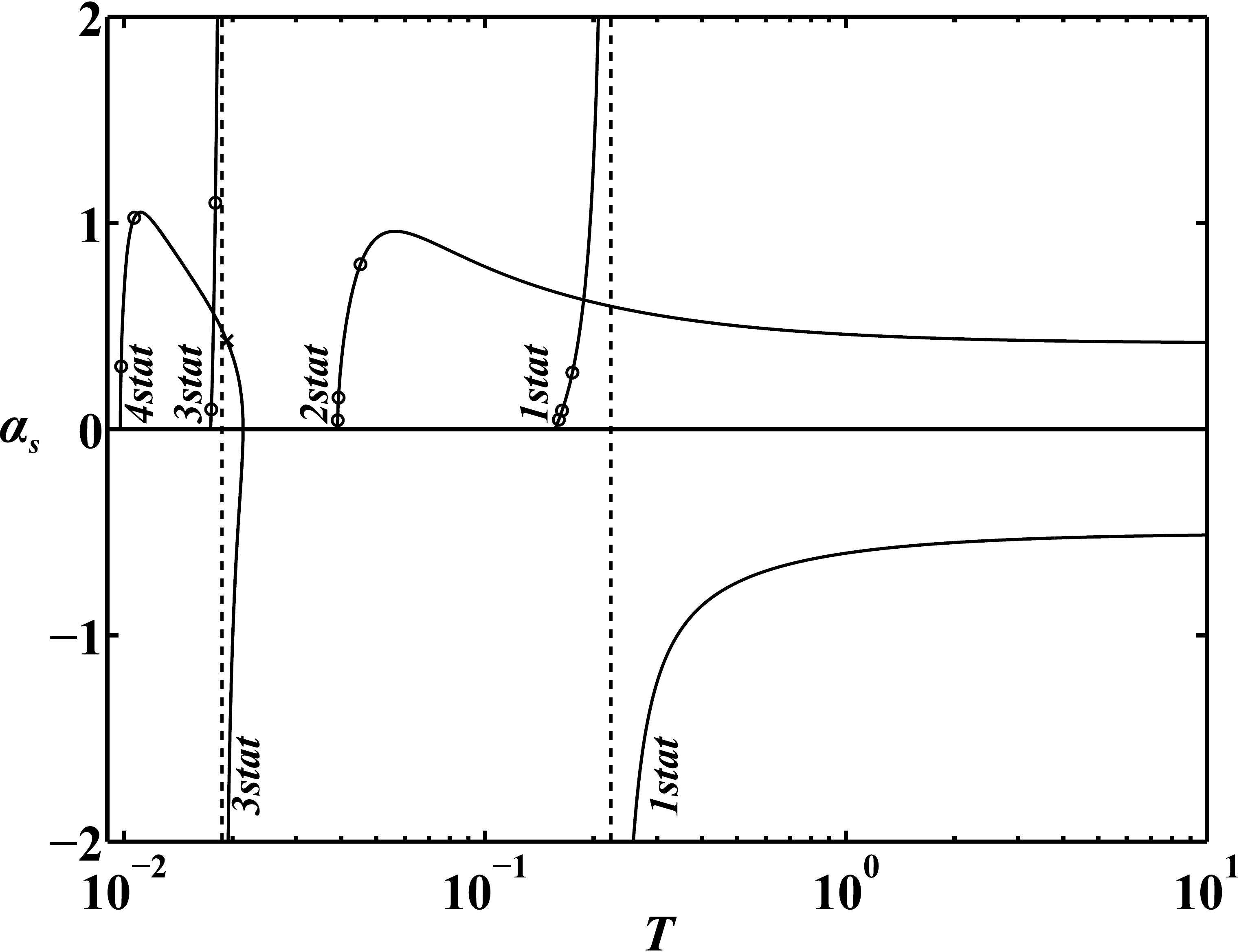}
			}\hfill
			\subfloat[\label{fig:aob_Stat}]{%
				\includegraphics[width=0.54\textwidth]{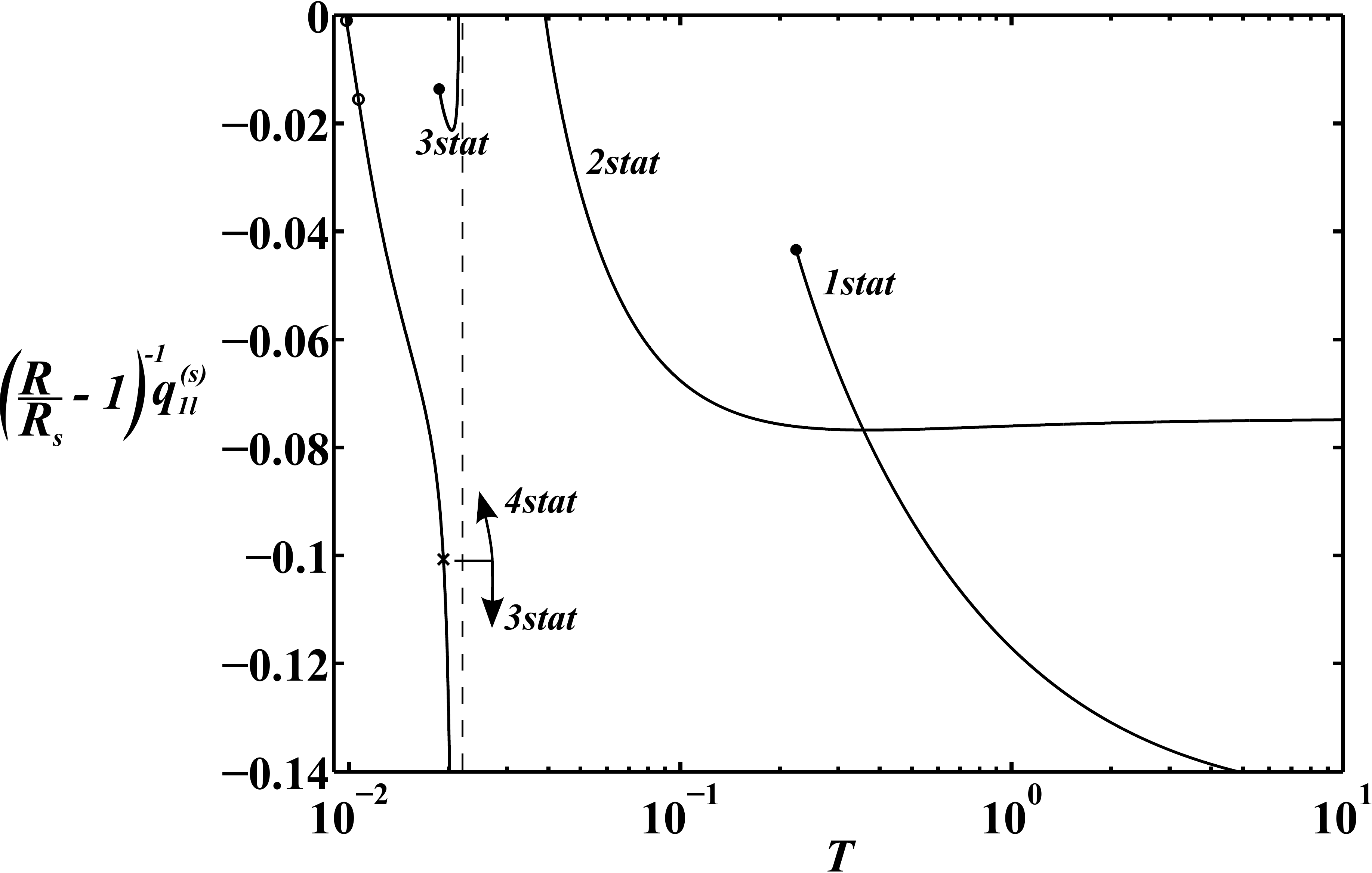}
			}
			\caption{Coefficients of the amplitude equation slightly above threshold versus $T$ for the steady bifurcating flow. (a) Growth rate factored by $(R/R_s - 1)^{-1}$. (b) Equilibrium amplitude multiplied by the same factor.}
			\label{fig:ab_Stat}
		\end{figure}

		Fig.\ \ref{fig:a_Stat} depicts the coefficient $\alpha_s$ (the growth rate scaled by $R/R_s - 1$) of the critical monotonic modes, as a function of $T$; it is important to keep in mind that instability corresponds to positive or negative values of $\alpha_s$, depending on the mode number. It is noticed that only modes with even mode number display a maximum whose value decreases with $T$. Close to Takens-Bogdanov points, the growth rate exhibits a singular behavior and changes sign on either side of those points. This \textit{a priori} has no physical sense since $\alpha_s$ must remain negative everywhere on the branch $1$stat for instability to occur. The coefficient $\alpha_s$ must again be negative on almost the entirety of the branches corresponding to critical modes of odd number. This anomaly is expected to be related to the complexity of the dynamics arising from interaction of static and oscillatory modes near Takens-Bogdanov points. A better understanding of the physics around such points requires a local analysis similar to that developed by \citeA{Xu13} in the high $R$ regime. This study is in due course but is beyond the scope of the present paper. Plots in Fig.\ \ref{fig:aob_Stat} represent the variations with $T$ of the eigenvalue $q_{1l}^{(s)}$ scaled by $(R/R_s - 1)$, it is proportional to the amplitude of the static membrane deformation when the equilibrium state is reached. Values of $T$ are limited only, as mentioned above, to those leading to positive values of growth rate $\alpha_s(R/R_s - 1)$. For those values of $T$, the product $(R/R_s - 1)^{-1}q_{1l}^{(s)}$ is always negative which means that the static divergence increases or decreases the mean flow rate, depending on the marginal mode number; it decreases rapidly with $T$ except for the mode $2$stat for which the rate of decrease is slower for sufficiently high values of $T$.
		
\begin{figure}[!h]
			\centering
			\subfloat[\label{fig:a_Osc}]{%
				\raisebox{-1.050\height}{\includegraphics[width=0.443\textwidth]{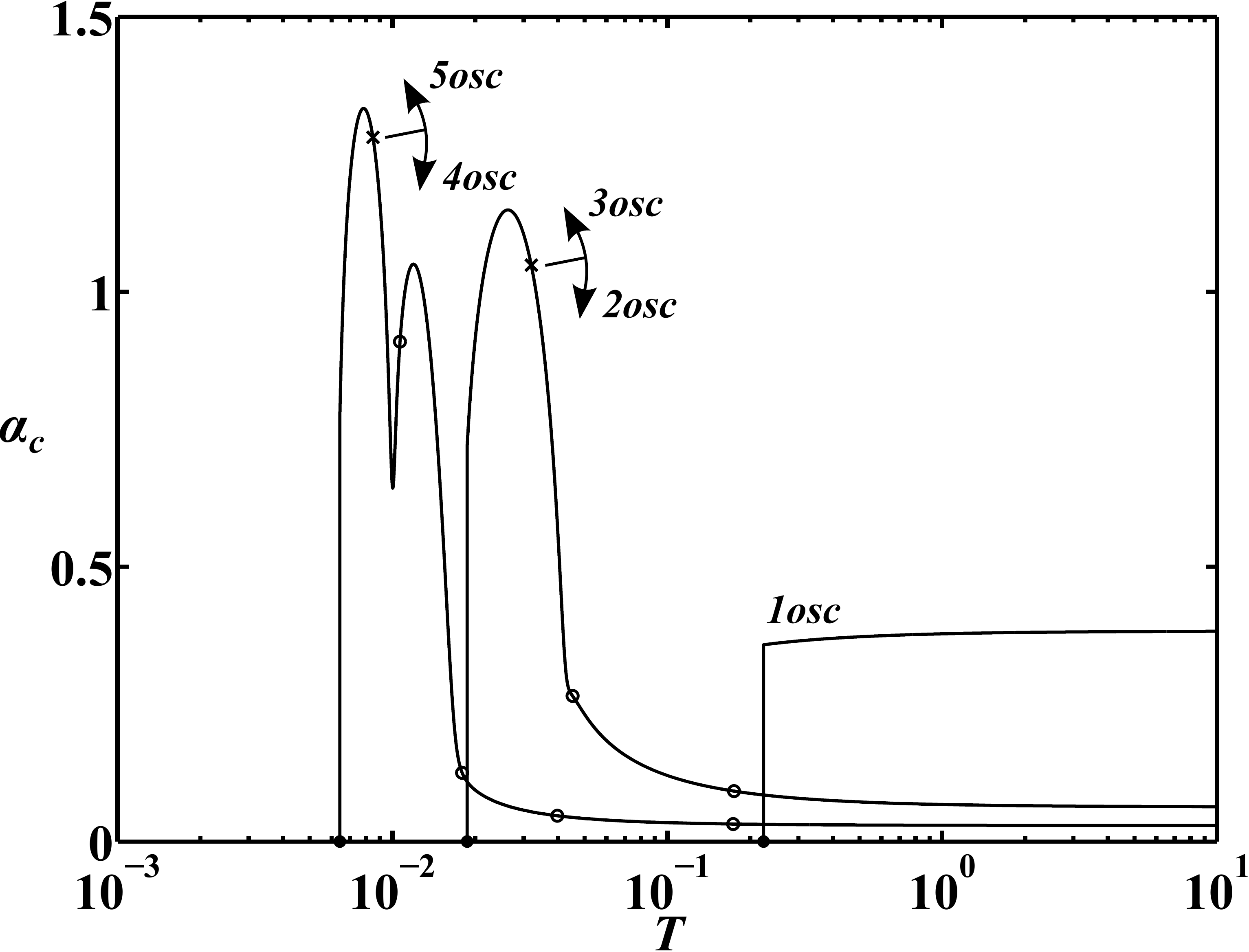}}
			}\hfill
			\subfloat[\label{fig:aob_Osc}]{%
				\includegraphics[width=0.514\textwidth]{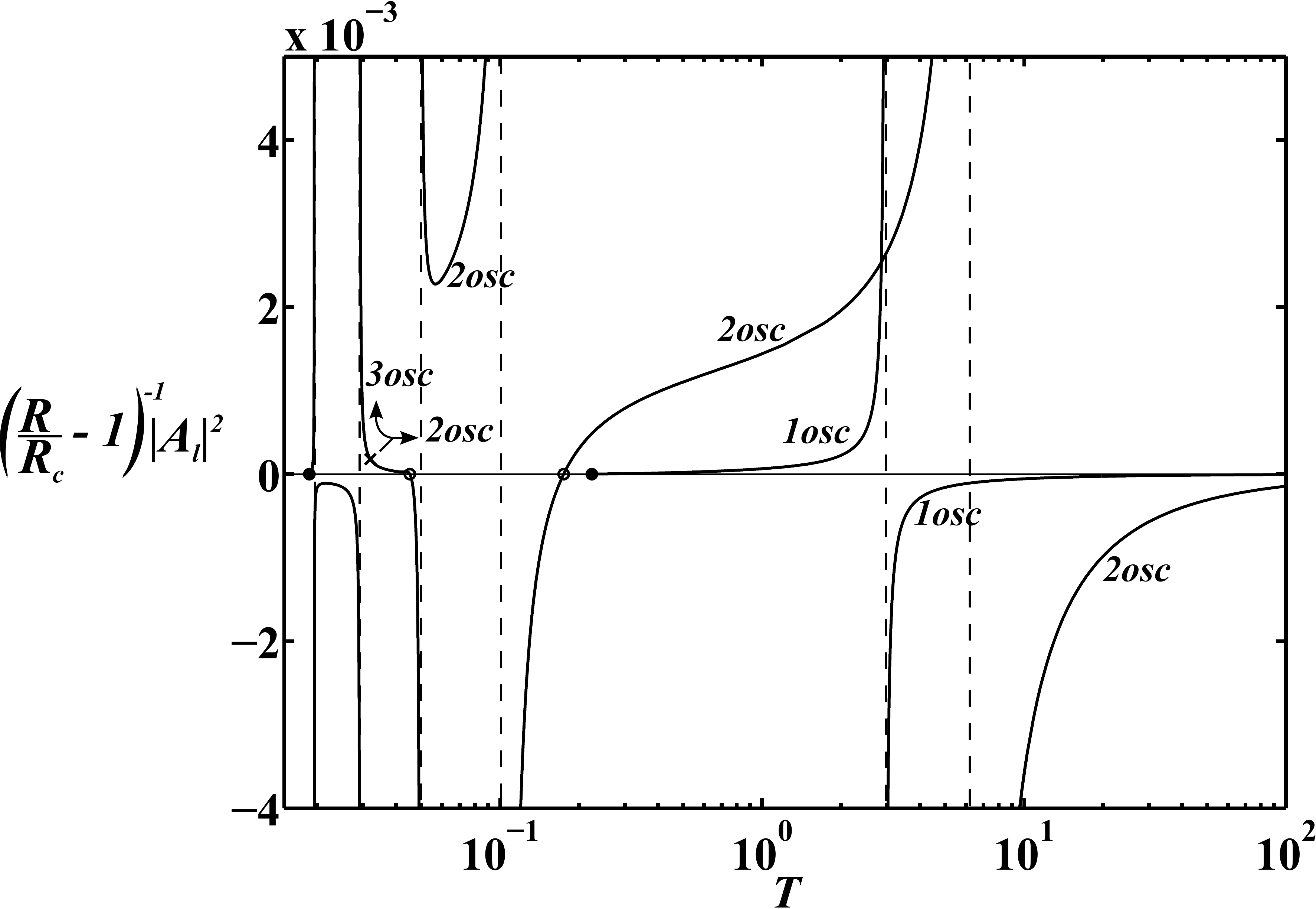}
			}
		\caption{Coefficients of the amplitude equation slightly above threshold versus $T$ for the oscillatory bifurcating flow. (a) Growth rate factored by $(R/R_c - 1)^{-1}$. (b) Equilibrium amplitude multiplied by the same factor.}
		\label{fig:ab_Osc}
		\end{figure}

		Fig.\ \ref{fig:a_Osc} depicts the growth rate $\alpha_c$, scaled by ($R/R_c - 1$), of the oscillatory modes against $T$. It can be observed that the degree of excitation follows the mode number; i.e., the higher the mode number, the larger the excitation, especially close to the Takens-Bogdanov points. Far from these points, the growth rate falls rapidly as $T$ increases for all modes except the mode $1$osc whose excitation remains almost constant (it is insensitive to variations of $T$). For this reason, the mode $1$osc is the dominant mode for values of $T$ greater than $0.2237$. The equilibrium amplitudes $A_l$ of the bifurcating oscillating modes, factored by $(R/R_c - 1)^{-1}$, are plotted versus $T$ in Fig.\ \ref{fig:aob_Osc}. Of utmost importance is the fact that the amplitudes of all the oscillating modes become singular at one or more particular values of $T$, namely, $T = 2.9530$ for $1$osc and $T = 0.0195$, $0.0291$, $0.0496$, $0.1052$, and $6.2218$ for $2$osc, for example. This breakdown is associated to the vanishing of the coefficients $\beta_c$, accounting for the cubic interaction of the modes when $R$ is slightly above $R_c$. This can be understood as a defect of the amplitude equations that might be alleviated by improving it through incorporating quintic order terms in view of the symmetry $A \rightarrow -A$. It is also worth noting that the product $(R/R_c - 1)^{-1}|A_l|^2$ is negative in some ranges of $T$. For these values of $T$, the cubic term does not provide saturation of the linear exponential growth. As emphasized above, quintic order terms should be added to complete the amplitude equations, leading then to subcritical bifurcations with hysteresis phenomena which is associated with bistability. Stable time-periodic solutions also exist in some range of $R$ below $R_c$ and can be reached either via finite amplitude perturbations or by starting from a time-periodic regime and decreasing $R$ below $R_c$ for a fixed $T$. The subcritical character of the Hopf bifurcations beyond some limit of $T$ is in accordance with the asymptotic result obtained by \citeA{Stewart09}.
		
		From a quantitative point of view, performing a similar expansion to the one developed in Appendix A of \citeA{Stewart09}, we found that the increase in the mean flowrate is $\beta$ times the one provided by the IM and the amplitude of the critical oscillatory mode is related to the one provided by the IM by a factor $\gamma^{-1/2}$. Finally, it is to be noted that at the linear and weakly nonlinear levels the differences between the WIM and the IM are only of quantitative character.
		
	\section{\label{sec:Conc}Summary}
	
		Linear and weakly nonlinear stability of planar Poiseuille flow in a channel with a flexible membrane as part of one of its walls is examined using a variant of the integral model. Despite the success of the IM in the nonlinear regime, it does not accurately predict neutral stability boundaries and introduces an error, typically of the order $20$\% (due primarily to the presence of the coefficient $\gamma$ in the model equations) in the critical values of $T$ and $R$. The present model corrects this discrepancy but does not introduce any qualitative change in the IM predictions at the linear level. A weakly nonlinear analysis has revealed the nature of the two bifurcations encountered in the explored parameter range. Hopf bifurcations are subcritical for sufficiently large values of $T$ for all oscillating modes; this is in conformity with the asymptotic calculation as $T \rightarrow \infty$. For values of $T$ smaller than about $6$, Hopf bifurcations are supercritical or subcritical depending on both $T$ and the mode number.
	
	\appendix
	\renewcommand{\theequation}{\thesection.\arabic{equation}}
	\renewcommand{\thefigure}{\thesection.\arabic{figure}}
	\renewcommand{\thetable}{\thesection.\arabic{table}}
		
	\section{\label{app:AdjProb1}Adjoint Problem \& Solvability Condition for (\ref{eq:Pmom_S2s}) subject to (\ref{eq:Pbcs_S2})}
	\setcounter{equation}{0}
	\setcounter{figure}{0}
	\setcounter{table}{0}
	
		The adjoint equation associated to Equation (\ref{eq:Pmom_S2s}) can be derived by premultiplying it by some function $h^{(a)}$ and transforming the integral $\int_0^1 h^{(a)}\cdot\mathcal{L}_sh_2^{(s)} \mathrm{d}x$ by repeated integration by parts and using the boundary condition $h_2^{(s)}(0) = h_2^{(s)}(1) = 0$, one obtains
		\begin{equation}
		\label{eq:A1}
			\int_0^1 h^{(a)}\cdot\mathcal{L}_sh_2^{(s)} \mathrm{d}x \equiv \int_0^1 h_2^{(s)}\cdot\mathcal{L}^{(a)}_sh^{(a)} \mathrm{d}x + Th_{2xx}^{(s)}h^{(a)}\Big{|}_0^1 - Th_{2x}^{(s)}h^{(a)}_x\Big{|}_0^1,
		\end{equation}
		where $\mathcal{L}^{(a)}_s$ is obtained from $\mathcal{L}_s$ by performing the change $\partial_x \rightarrow -\partial_x$. Now we define the adjoint operator $\mathcal{L}_s^{(a)}$ associated to the original problem (\ref{eq:Pcont_S2})-(\ref{eq:Pbcs_S2}) by $\mathcal{L}^{(a)}_sh^{(a)} = 0$, $h_x^{(a)}(1) = h_x^{(a)}(0) = 0$ which gives the adjoint eigenfunction
		\begin{equation}
		\label{eq:A2}
			h^{(a)}(x) = c_0\left(ce^{\lambda x} + \bar{c}e^{\bar{\lambda}x} + e^{-(\lambda + \bar{\lambda})x}\right),
		\end{equation}
		where the boundary conditions $h_x^{(a)}(1) = h_x^{(a)}(0) = 0$ imply
		$$
			c = \frac{(\lambda + \bar{\lambda})\left(1 - e^{-(\lambda + 2\bar{\lambda})}\right)}{\lambda\left(1 - e^{(\lambda - \bar{\lambda})}\right)}
		$$
		and $c_0$ is a nonzero real constant which can be set equal to unity. Equation (\ref{eq:A1}) then takes the reduced form
		\begin{equation}
		\label{eq:A3}
			\int_0^1h^{(a)}(x)S(x)\mathrm{d}x + \left[Th_{2xx}^{(s)}h^{(a)}\right]^0_1 = 0
		\end{equation}
		which expresses the orthogonality of $h^{(a)}$ and the source terms in (\ref{eq:Pmom_S2}) and (\ref{eq:Pbcs_S2}). Note that only two boundary conditions are sufficient to define the adjoint problem because $R_s$ is also an eigenvalue of $\mathcal{L}_s^{(a)}$ corresponding to $h^{(a)}$. It is possible to define the adjoint problem in another manner, namely by imposing a third boundary condition (a linear relation between $h^{(a)}(1)$ and $h^{(a)}(0)$, for example). This would lead to another spectrum of the adjoint operator and a different adjoint solution. A third boundary condition is however not required in the present case since $Th_{2xx}\Big{|}_{0,1}$ is known at the boundaries.
		
	\section{\label{app:Source}Expression of Source Terms in (\ref{eq:Psource_O3})}
	\setcounter{equation}{0}
	\setcounter{figure}{0}
	\setcounter{table}{0}
	
		$$
			\begin{array}{l}
				S_{31}(x) = -i\gamma{\omega}Q_1, \\
				S_{32}(x) = \displaystyle\frac{36}{R_c}Q_{1x} + \displaystyle\frac{12}{R_c}i{\omega}Q_1, \\
				\begin{aligned}
					& S_{33}(x) = -\left(\frac{12}{5}{\alpha}Q_{1xx} + \frac{12}{5}i\beta{\omega}Q_{1x}\right)Q_{20} - 3TQ_{1x}H_{20xxx} + \frac{12}{5}i\alpha{\omega}Q_1H_{20x} \\ &
					-\left[3TQ_{1xxxx} + \left(\frac{72}{R_c} + \frac{12}{5}i\beta\omega\right)Q_{1x} - 2\gamma\omega^2Q_1\right]H_{20} - 3T\bar{Q}_{1x}Q_{22xxxx} \\ &
					- \frac{12}{5}i\alpha\omega\bar{Q}_1Q_{22xx} - \left[3T\bar{Q}_{1xxxx} + \left(\frac{72}{R_c} + \frac{12}{5}i\beta\omega\right)\bar{Q}_{1x} - \left(\frac{24}{5}\beta\omega^2 + 2\gamma\omega^2\right)\bar{Q}_1\right]Q_{22x} \\ &
					+ \left[\frac{24}{5}i\alpha\omega\bar{Q}_{1xx} + \left(\frac{24}{5}\beta\omega^2 + 8 \gamma\omega^2\right)\bar{Q}_{1x}\right]Q_{22} - \left[3T\bar{Q}_{1xxxx} - \left(\frac{12}{5}\beta\omega^2 + \gamma\omega^2\right)\bar{Q}_1\right]Q_{1x}^2 \\ &
					- \left(6TQ_{1xxxx} + \frac{36}{R_c}Q_{1x} - 2\gamma\omega^2Q_1\right)|Q_{1x}|^2 + \frac{6}{5}\alpha\omega^2\bar{Q}_{1xx}Q_1^2 - \frac{12}{5}\alpha\omega^2Q_{1xx}|Q_1|^2.
				\end{aligned}
			\end{array}
		$$
	
	\section{\label{app:AdjProb2}Adjoint Problem \& Solvability Condition for (\ref{eq:Pmom_O3s}) subject to (\ref{eq:Pbcs_O3})}
	\setcounter{equation}{0}
	\setcounter{figure}{0}
	\setcounter{table}{0}
	
		In order to derive a solvability condition for the problem (\ref{eq:Pmom_O3s})-(\ref{eq:Pbcs_O3}), we must define a suitable scalar product and the associated adjoint problem. We first define the adjoint operator of $\mathcal{L}_\omega$ as $\mathcal{L}_\omega^{(a)} = \bar{\mathcal{L}}_\omega^*$ with $\mathcal{L}_\omega^*$ being obtained from $\mathcal{L}_\omega$ by the change $\partial_x \rightarrow -\partial_x$. Let $Q_a$ be the eigenfunction of $\mathcal{L}_\omega^{(a)}$, i.e.,
		\begin{equation}
		\label{eq:C1}
			\mathcal{L}_\omega^{(a)}Q_a = 0
		\end{equation}
		satisfying the following boundary conditions:
		\begin{equation}
		\label{eq:C2}
			\begin{array}{c}
				Q_{ax}\Big{|}_0 = Q_{ax}\Big{|}_1 = 0, \\[4pt]
				TQ_{axxx}\Big{|}_0 = (\bar{A}_0 + \bar{B})Q_a(0), \\[4pt]
				TQ_{axxx}\Big{|}_1 = (\bar{A}_1 + \bar{B})Q_a(1),
			\end{array}
		\end{equation}
		where
		$$
			A_0 = L_1\left(\gamma\omega^2 - \frac{12}{R_c}i\omega\right), \quad A_1 = -L_2\left(\gamma\omega^2 - \frac{12}{R_c}i\omega\right), \quad \text{and} \quad B = \frac{36}{R_c} + \frac{12}{5}i\beta\omega
		$$
		are the factors of the first $x$ derivative in $\mathcal{L}_\omega$. Then upon multiplying (\ref{eq:Pmom_O3s}) by $\bar{Q}_a$ and performing successive integrations by parts yields the solvability condition
		\begin{equation}
		\label{eq:C3}
			\int_0^1 \tilde{S}_3(x)\cdot\bar{Q}_a \mathrm{d}x + B_0\bar{Q}_a(0) + B_1\bar{Q}_a(1) = 0
		\end{equation}
		which expresses the orthogonality of the source term $(S_3(x),B_0,B_1)$ and the complex conjugate of the spanning vector of the kernel of the adjoint operator $\mathcal{L}_\omega^{(a)}$.
		
	\bibliographystyle{apacite}

\end{document}